\documentclass[11pt,a4paper, DIV12]{scrartcl}

\usepackage[T1]{fontenc}
\usepackage[utf8]{inputenc}

\usepackage{lineno}  %
\usepackage{xspace} %
\usepackage{setspace}
\usepackage[format=hang,
			textfont={small,it},
            font={small,it},
			labelsep=quad]{caption}

\usepackage{multirow}

\usepackage[symbol*,perpage]{footmisc}

\usepackage{graphicx}  %
\usepackage{color}
\usepackage{colortbl}
\usepackage{multirow}

\usepackage{amsmath} %
\usepackage{bm}
\usepackage{amssymb}
\usepackage{amsfonts}
\usepackage{upgreek} %

\usepackage{booktabs} %

\usepackage{appendix}

\usepackage{numprint}

\usepackage[]{tocstyle}
\usetocstyle{nopagecolumn}

\usepackage[affil-it]{authblk} 

\usepackage{ifthen} %
 
\newboolean{uprightparticles}
\setboolean{uprightparticles}{false} %

\usepackage{xspace} 
\usepackage{upgreek}

\def\lhcb {\mbox{LHCb}\xspace}

\def\belle  {\mbox{Belle}\xspace}

\def\lhc    {\mbox{LHC}\xspace}

\def\MagUp {\mbox{\em Mag\kern -0.05em Up}\xspace}

\ifthenelse{\boolean{uprightparticles}}%
{
 
 \def\Pgamma      {\ensuremath{\upgamma}\xspace}

 \def\Ppi         {\ensuremath{\uppi}\xspace}

 \def\PDelta      {\ensuremath{\Delta}\xspace}                 
 \def\PXi      {\ensuremath{\Xi}\xspace}                 
 \def\PLambda      {\ensuremath{\Lambda}\xspace}                 
 \def\PSigma      {\ensuremath{\Sigma}\xspace}                 
 \def\POmega      {\ensuremath{\Omega}\xspace}                 
 \def\PUpsilon      {\ensuremath{\Upsilon}\xspace}

 \def\PB      {\ensuremath{\mathrm{B}}\xspace}                 
                  
 \def\PD      {\ensuremath{\mathrm{D}}\xspace}

 \def\PH      {\ensuremath{\mathrm{H}}\xspace}

 \def\PK      {\ensuremath{\mathrm{K}}\xspace}

 \def\PW      {\ensuremath{\mathrm{W}}\xspace}

 \def\PZ      {\ensuremath{\mathrm{Z}}\xspace}                 
                  
 \def\Pb      {\ensuremath{\mathrm{b}}\xspace}

 \def\Pe      {\ensuremath{\mathrm{e}}\xspace}

 \def\Pi      {\ensuremath{\mathrm{i}}\xspace}

 \def\Ps      {\ensuremath{\mathrm{s}}\xspace}

}
{
 
 \def\Pgamma      {\ensuremath{\gamma}\xspace}

 \def\Ppi         {\ensuremath{\pi}\xspace}

 \mathchardef\PDelta="7101
 \mathchardef\PXi="7104
 \mathchardef\PLambda="7103
 \mathchardef\PSigma="7106
 \mathchardef\POmega="710A
 \mathchardef\PUpsilon="7107
                  
 \def\PB      {\ensuremath{B}\xspace}                 
                  
 \def\PD      {\ensuremath{D}\xspace}

 \def\PH      {\ensuremath{H}\xspace}

 \def\PK      {\ensuremath{K}\xspace}

 \def\PW      {\ensuremath{W}\xspace}

 \def\PZ      {\ensuremath{Z}\xspace}                 
                  
 \def\Pb      {\ensuremath{b}\xspace}

 \def\Pe      {\ensuremath{e}\xspace}

 \def\Pi      {\ensuremath{i}\xspace}

 \def\Ps      {\ensuremath{s}\xspace}

}

\makeatletter
\ifcase \@ptsize \relax%
  \newcommand{\miniscule}{\@setfontsize\miniscule{4}{5}}%
\or%
  \newcommand{\miniscule}{\@setfontsize\miniscule{5}{6}}%
\or%
  \newcommand{\miniscule}{\@setfontsize\miniscule{5}{6}}%
\fi
\makeatother

\DeclareRobustCommand{\optbar}[1]{\shortstack{{\miniscule (\rule[.5ex]{1.25em}{.18mm})}
  \\ [-.7ex] $#1$}}

\def\epem       {{\ensuremath{\Pe^+\Pe^-}}\xspace}

\def\g      {{\ensuremath{\Pgamma}}\xspace}
\def\H      {{\ensuremath{\PH^0}}\xspace}

\def\W      {{\ensuremath{\PW}}\xspace}

\def\Z      {{\ensuremath{\PZ}}\xspace}

\def\squark    {{\ensuremath{\Ps}}\xspace}

\def\bquark    {{\ensuremath{\Pb}}\xspace}

\def\pion   {{\ensuremath{\Ppi}}\xspace}
\def\piz    {{\ensuremath{\pion^0}}\xspace}

\def\pip    {{\ensuremath{\pion^+}}\xspace}
\def\pim    {{\ensuremath{\pion^-}}\xspace}

\def\kaon    {{\ensuremath{\PK}}\xspace}
  \def\Kbar    {{\kern 0.2em\overline{\kern -0.2em \PK}{}}\xspace}

\def\KorKbar    {\kern 0.18em\optbar{\kern -0.18em K}{}\xspace}

\def\Kp      {{\ensuremath{\kaon^+}}\xspace}

\def\Kstarz  {{\ensuremath{\kaon^{*0}}}\xspace}

\def\Kstar   {{\ensuremath{\kaon^*}}\xspace}

  \def\Dbar    {{\kern 0.2em\overline{\kern -0.2em \PD}{}}\xspace}
\def\D       {{\ensuremath{\PD}}\xspace}

\def\DorDbar    {\kern 0.18em\optbar{\kern -0.18em D}{}\xspace}

\def\B       {{\ensuremath{\PB}}\xspace}
\def\Bbar    {{\ensuremath{\kern 0.18em\overline{\kern -0.18em \PB}{}}}\xspace}

\def\BorBbar    {\kern 0.18em\optbar{\kern -0.18em B}{}\xspace}
\def\Bz      {{\ensuremath{\B^0}}\xspace}

\def\Bu      {{\ensuremath{\B^+}}\xspace}

\def\Bp      {{\ensuremath{\Bu}}\xspace}

\def\Bs      {{\ensuremath{\B^0_\squark}}\xspace}

  \def\Y#1S{\ensuremath{\PUpsilon{(#1S)}}\xspace}%

\def\Lbar        {{\ensuremath{\kern 0.1em\overline{\kern -0.1em\PLambda}}}\xspace}
\def\LorLbar    {\kern 0.18em\optbar{\kern -0.18em \PLambda}{}\xspace}

\newcommand{\decay}[2]{\ensuremath{#1\!\to #2}\xspace}         %

\def\to                 {\ensuremath{\rightarrow}\xspace}

\def\CP                {{\ensuremath{C\!P}}\xspace}

\def\AT#1     {\ensuremath{A_{\mathrm{T}}^{#1}}\xspace}           %
\def\btosgam  {\decay{\bquark}{\squark \g}}

\def\C#1      {\ensuremath{\mathcal{C}_{#1}}\xspace}                       %
\def\Cp#1     {\ensuremath{\mathcal{C}_{#1}^{'}}\xspace}                    %
\def\Ceff#1   {\ensuremath{\mathcal{C}_{#1}^{\mathrm{(eff)}}}\xspace}        %
\def\Cpeff#1  {\ensuremath{\mathcal{C}_{#1}^{'\mathrm{(eff)}}}\xspace}       %
\def\Ope#1    {\ensuremath{\mathcal{O}_{#1}}\xspace}                       %
\def\Opep#1   {\ensuremath{\mathcal{O}_{#1}^{'}}\xspace}                    %

\newcommand{\tev}{\ifthenelse{\boolean{inbibliography}}{\ensuremath{~T\kern -0.05em eV}}{\ensuremath{\mathrm{\,Te\kern -0.1em V}}}\xspace}
\newcommand{\gev}{\ensuremath{\mathrm{\,Ge\kern -0.1em V}}\xspace}
\newcommand{\mev}{\ensuremath{\mathrm{\,Me\kern -0.1em V}}\xspace}
\newcommand{\kev}{\ensuremath{\mathrm{\,ke\kern -0.1em V}}\xspace}
\newcommand{\ev}{\ensuremath{\mathrm{\,e\kern -0.1em V}}\xspace}
\newcommand{\gevc}{\ensuremath{{\mathrm{\,Ge\kern -0.1em V\!/}c}}\xspace}
\newcommand{\mevc}{\ensuremath{{\mathrm{\,Me\kern -0.1em V\!/}c}}\xspace}
\newcommand{\gevcc}{\ensuremath{{\mathrm{\,Ge\kern -0.1em V\!/}c^2}}\xspace}
\newcommand{\gevgevcccc}{\ensuremath{{\mathrm{\,Ge\kern -0.1em V^2\!/}c^4}}\xspace}
\newcommand{\mevcc}{\ensuremath{{\mathrm{\,Me\kern -0.1em V\!/}c^2}}\xspace}

\def\m    {\ensuremath{\mathrm{ \,m}}\xspace}

\def\invfb   {\ensuremath{\mbox{\,fb}^{-1}}\xspace}

\def\invab   {\ensuremath{\mbox{\,ab}^{-1}}\xspace}

\def\gsim{{~\raise.15em\hbox{$>$}\kern-.85em
          \lower.35em\hbox{$\sim$}~}\xspace}
\def\lsim{{~\raise.15em\hbox{$<$}\kern-.85em
          \lower.35em\hbox{$\sim$}~}\xspace}

\def\tell1  {TELL1\xspace}
\def\ukl1   {UKL1\xspace}

\newcommand{\ie}{\mbox{\itshape i.e.}\xspace}
\newcommand{\etal}{\mbox{\itshape et al.}\xspace}

\newcommand{\diff}[1]{\operatorname{d}\!#1}

\def\Kres{\ensuremath{K_{\text{res}}}\xspace}

\def\BKresgammaKpipigamma{\ensuremath{\decay{\B}{\decay{\Kres\gamma}{\kaon\pion\pion\g}}}\xspace}

\def \Kres{\ensuremath{\kaon_{\text{res}}}\xspace}
\def \Kresp{\ensuremath{\kaon^{i+}_{\text{res}}}\xspace}

\def \Kpipi{\ensuremath{\kaon\pion\pion}\xspace}

\def \Kpipig{\ensuremath{\kaon\pion\pion\g}\xspace}

\def \BtoKpipig{\decay{B}{ K\pi\pi\g}}

\def \BptoKpipig{\decay{\Bp}{\Kp\pim\pip\g}}
\def \BztoKpipizg{\decay{\Bz}{\Kp\pim\piz\g}}

\def \BptoKresg{\ensuremath{\decay{\Bp}{\Kresp{\g}}}\xspace}

\def \Aud{\ensuremath{\mathcal{A}_{\text{ud}}}\xspace}
\def \lg{\ensuremath{\lambda_{\gamma}}\xspace}

\usepackage[backend=biber,
            style=phys,
            eprint=true,
            hyperref=true,
            biblabel=brackets,
            maxcitenames=5,
            maxbibnames=5,
            bibencoding=utf8,
            url=true,
            defernumbers=false]{biblatex}
\DefineBibliographyStrings{english}{%
  andothers = {\etal} %
}
\DeclareFieldFormat*{title}{\textit{#1}} %

\DeclareSourcemap{
  \maps[datatype=bibtex,overwrite=true]{
    \map{
      \step[fieldset=eprintclass,fieldvalue=] %
    }
    \map{
      \step[fieldsource=collaboration,final=true] %
      \step[fieldset=usera,origfieldval,final=true]
    }
  }
}
\renewbibmacro*{author}{%
  \iffieldundef{usera}{%
    \printnames{author} %
  }{%
    \printfield{usera}, \printnames{author} %
  }
}

\usepackage{hyperref}

\title{\vspace{10mm}\huge Using an amplitude analysis\\to measure the photon polarisation\\in $\bm{B \rightarrow K\pi\pi\gamma}$ decays\vspace{20mm}}

\author[1]{V.~Bell\'{e}e}
\author[1]{P.~Pais}
\author[2]{A.~Puig~Navarro}
\author[1]{F.~Blanc}
\author[1]{O.~Schneider}
\author[3]{K.~Trabelsi}
\author[3$*$]{G.~Veneziano}

\affil[1]{\footnotesize Institute of Physics, Ecole Polytechnique  F{\'e}d{\'e}rale de Lausanne (EPFL), Lausanne, Switzerland}
\affil[2]{\footnotesize Physik-Institut, Universit{\"a}t Z{\"u}rich, Z{\"u}rich, Switzerland}
\affil[3]{\footnotesize LAL, Univ. Paris-Sud, CNRS/IN2P3, Universit{\'e} Paris-Saclay, Orsay, France}
\affil[$*$]{\footnotesize Affiliation at the time this work was carried out}

\date{\small{\today}}

\begin{document}

\maketitle

\vspace{8mm}

\begin{abstract}
\begin{center}
\textbf{Abstract}
\end{center}

\noindent A method is proposed to measure the photon polarisation parameter \lg in \btosgam transitions using an amplitude analysis of \BtoKpipig decays.
Simplified models of the \Kpipi system are used to simulate \BptoKpipig and \BztoKpipizg decays, validate the amplitude analysis method, and demonstrate the feasibility of a measurement of the \lg parameter irrespective of the model parameters.
Similar sensitivities to \lg are obtained with both the charged and neutral hadronic systems.
In the absence of any background and distortion due to experimental effects, the statistical uncertainty expected from an analysis of \BptoKpipig decays in an \lhcb data set corresponding to an integrated luminosity of $9$~\invfb is estimated to be $0.009$.
A similar measurement using \BztoKpipizg decays in a Belle II data sample corresponding to an integrated luminosity of $5$~\invab would lead to a statistical uncertainty of $0.018$.

\vspace{15mm}

\end{abstract}

\setcounter{tocdepth}{1}

\newpage

\clearpage

\section{Introduction\label{sec:intro}}

Rare \btosgam flavour-changing neutral-current transitions are expected to be sensitive to New Physics (NP) effects. 
These transitions are allowed only at loop level, and NP could arise from the exchange of a heavy particle in the electroweak penguin loop. 
In the Standard Model (SM), the recoil \squark quark that couples to a \W boson is left-handed, causing the photon emitted in \btosgam transitions to be almost completely left-handed. 
Several theories beyond the SM predict a significant right-handed component for the photon polarisation: in the minimal supersymmetric model (MSSM), left-right squark mixing causes a chirality flip along the gluino line in the electroweak penguin loop~\cite{Everett:2001yy}, while in some grand unification models right-handed neutrinos (and the associated right-handed quark coupling) are expected to enhance the right-handed photon component~\cite{becirevic2012}.

Various complementary approaches have been proposed for the determination of the polarisation of the photon in \btosgam transitions. 
An indirect method consists in studying the time-dependent decay rate of $\B^{0}_{(\squark)} \to f^{\CP} \gamma$ decays, where $f^{\CP}$ is a particle or system of particles in a \CP eigenstate~\cite{Muheim2008174}.
An alternative approach involves the study of angular distributions of the four-body final state in $\Bz \to K^{*0} \ell^+ \ell^-$ decays \cite{Kruger:2005ep}.
Yet another proposed method involves exploiting the angular distributions of the photon and the proton in the final state of $\Lambda_\bquark \to \Lambda_X (\to p h) \gamma$ decays, where $\Lambda_X$ is either the ground state or an excited state of the $\Lambda$ hyperon and $h$ is a kaon or a pion~\cite{Hiller:2007ur}. 

Information on the photon polarisation can also be obtained from \PB decays to three hadrons and a photon.
This approach is enabled by the fact that the three final-state hadrons allow the construction of a parity-odd triple product that inverts its sign with a change in the photon chirality, and by the existence of interference between the amplitudes of the hadronic system.

In $B \to \Kres \gamma$ decays, where \Kres is a kaonic resonance decaying to a \Kpipi final state, the required interference in the \Kpipi system can arise from several sources. 
In the case of a single \Kres state, the helicity amplitudes must contain at least two terms with a non-vanishing relative phase. This can occur between intermediate resonance amplitudes in the decay \Kres $\rightarrow$ \Kpipi, between $S$ and $D$ wave amplitudes in the decay, or between two intermediate \Kstar\pion states with different charges, related by isospin symmetry.\footnote{This last type of interference is possible only in decays containing a \piz in the final state.}
Interference can also appear in the presence of different overlapping \Kres states;
in fact, the presence of a multitude of interfering resonances makes it very difficult to distinguish them, thus complicating the interpretation of the observed distributions.

A simplified approach to the study of the photon polarisation consists in exploiting the distribution of the polar angle of the photon with respect to the hadronic decay plane integrating over the resonance content of the \Kpipi system~\cite{Gronau:2001ng}.
Using $3\,\invfb$ of $pp$ collisions at the \lhc, the \lhcb collaboration determined the shape of this distribution and the \emph{up-down asymmetry} between the number of events with photons emitted on either side of the plane~\cite{LHCb-PAPER-2014-001}.
The up-down asymmetry was found to differ from zero by $5.2$ standard deviations. 
As this asymmetry is expected to be proportional to the photon polarisation parameter $\lambda_{\gamma}$, this result represents the first observation of a parity-violating nonzero photon polarisation in \btosgam transitions.
The proportionality coefficient between the up-down asymmetry and \lg depends on the resonance content of the \Kpipi system, and in particular on the interference pattern between the various decay modes. 
Without precise knowledge of these amplitudes, a measurement of the up-down asymmetry cannot be translated into a photon polarisation value.

In this paper, a method to determine the value of the photon polarisation parameter by means of an amplitude analysis of the \Kpipig system is proposed.
It is organised as follows: a description of the up-down asymmetry and its limitations in extracting a value for the photon polarisation parameter are detailed in Sec.~\ref{sec:motivation}.
In Sec.~\ref{sec:method}, a general expression for the \BtoKpipig decay rate in terms of a photon polarisation parameter is derived, the amplitude formalism is described, and the fit method used for the amplitude analysis is explained.
In Sec.~\ref{sec:sensitivity}, results for simulated data sets with assumed models of \BptoKpipig and \BztoKpipizg decays are presented. 
Statistical sensitivities on the photon polarisation parameter are quoted for these models, assuming no background and no experimental effect.
Conclusions are drawn in Sec.~\ref{sec:conclusions}.

\section{Motivation \label{sec:motivation}}

\BptoKpipig and \BztoKpipizg decays can be described in terms of five independent variables: two angles (cos$\,\theta$ and $\chi$) that describe the direction of the photon in the rest frame of the kaonic resonance $K_{\text{res}}$, and three squared invariant masses ($s_{123}, s_{12}, s_{23}$), where the indices $1$, $2$ and $3$ refer respectively to the final-state $\pi^+$, $\pi^-$ and $K^+$ for the charged decay mode, and to $\pi^-$, $\pi^0$ and $K^+$ for the neutral decay mode.

\begin{figure}[tb]
\centering
          \includegraphics[width=0.55\textwidth]{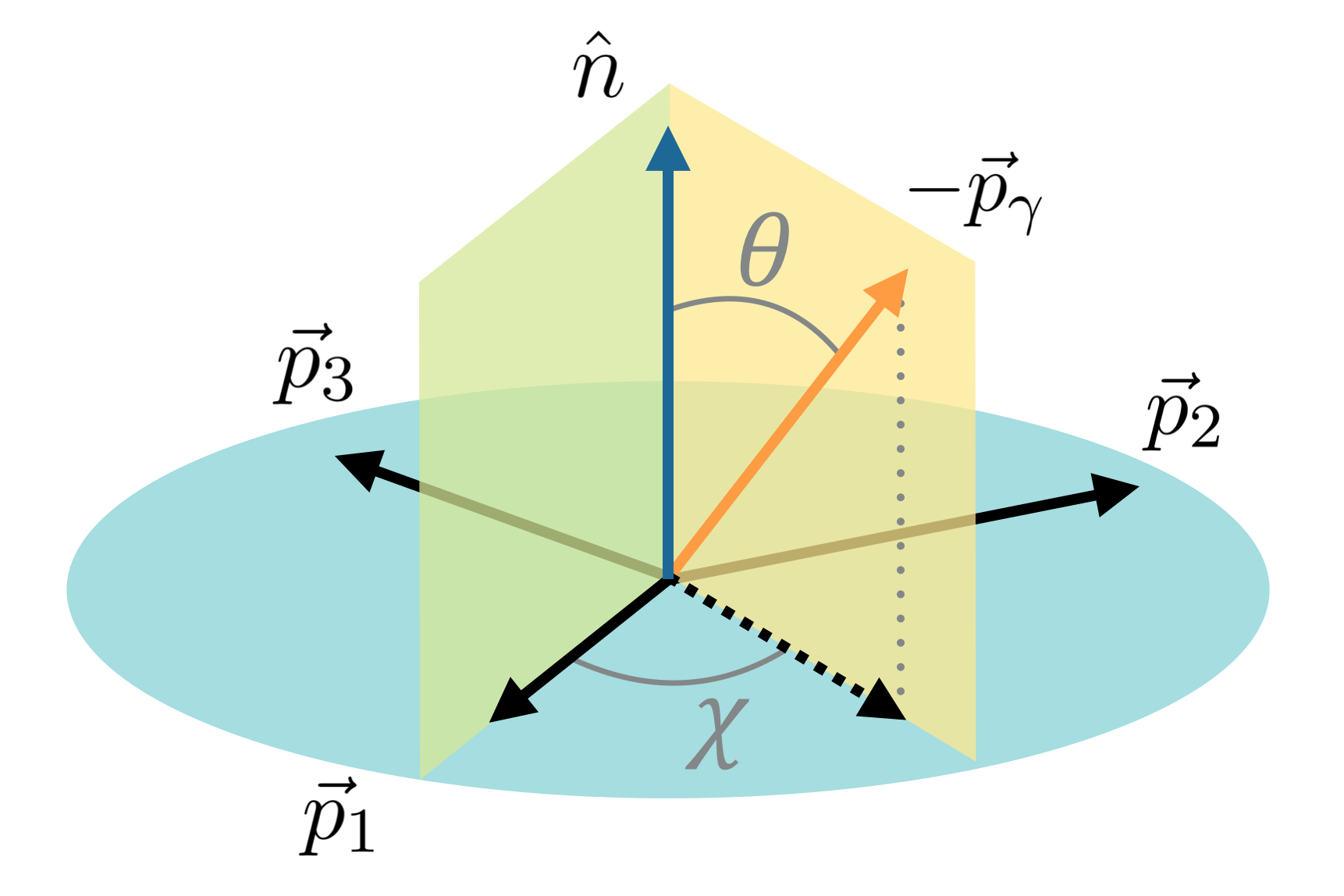}
\caption{Definitions of the angular variables used to describe the $K \pi \pi \gamma$ system. The indices $1$, $2$ and $3$ refer respectively to the final-state $\pi^+$, $\pi^-$ and $K^+$ in \BptoKpipig decays, and to $\pi^-$, $\pi^0$ and $K^+$ in \BztoKpipizg decays.\label{fig:def_costheta_chi}}

\end{figure}

As illustrated in Fig.~\ref{fig:def_costheta_chi} for \BptoKpipig decays, in the rest frame of the kaonic resonance $K_{\text{res}}$, the normal to the hadronic decay plane is denoted by $\hat{n} = (\vec{p}_{1} \times \vec{p}_{2}) / |\vec{p}_{1} \times \vec{p}_{2}|$. 
The polar angle $\theta$ is the angle between $\hat{n}$ and the opposite of the photon momentum, so that $\text{cos}\,\theta = - \hat{n}\cdot \vec{p}_{\gamma}/ |\vec{p}_{\gamma}| $\footnote{This definition of the polar angle corresponds to the one used in Ref.~\cite{Gronau:2002rz} and does not match the one in Ref.~\cite{LHCb-PAPER-2014-001}.}.
The angle $\chi$ is defined from

\begin{align}
\text{cos}\,\chi &= \frac{(\hat{n} \times \vec{p}_{1}) \cdot (\hat{n} \times \vec{p}_{\gamma})}{|\hat{n} \times \vec{p}_{1}| \, |\hat{n} \times \vec{p}_{\gamma}|}\,,  \\
\text{sin}\,\chi &= \frac{(\hat{n} \times \vec{p}_{1}) \times (\hat{n} \times \vec{p}_{\gamma})}{|\hat{n} \times \vec{p}_{1}| \, |\hat{n} \times \vec{p}_{\gamma}|} \cdot \hat{n}\,.
\end{align}

The \BtoKpipig differential branching fraction has the following dependence on $\cos\theta$~\cite{Gronau:2002rz}:

\begin{align}\label{eq:diffRate}
    \frac{\diff{\Gamma(\BKresgammaKpipigamma)}}{\diff{s_{123}}\diff{s_{12}}\diff{s_{23}}\diff{\chi}\diff{\cos\theta}} &= \sum_{i=0,2,4} a_i(s_{123}, s_{12}, s_{23},\chi)\cos^i\theta \nonumber \\
    &+ \lambda_\gamma \sum_{j=1,3} a_j(s_{123}, s_{12}, s_{23}, \chi)\cos^j\theta\,.
\end{align}

Integrating Eq.~\ref{eq:diffRate} over the squared invariant masses and $\chi$, the \emph{up-down} asymmetry (\Aud) is defined as~\cite{Gronau:2001ng,Gronau:2002rz}

\begin{equation}\label{eq:updown}
    \mathcal{A}_{\text{ud}} \equiv \frac{\int_0^1\diff{\cos{\theta}}\frac{\diff{\Gamma}}{\diff{\cos{\theta}}}-\int_{-1}^0\diff{\cos{\theta}}\frac{\diff{\Gamma}}{\diff{\cos{\theta}}}}{\int_{-1}^1\diff{\cos{\theta}}\frac{\diff{\Gamma}}{\diff{\cos{\theta}}}}, 
\end{equation}
where the terms in even powers of $\cos{\theta}$ disappear, and the resulting asymmetry is directly proportional to $\lambda_\gamma$ with a proportionality coefficient that depends on the resonance content of the \Kpipi system.

\begin{figure}[tb]
\centering
\includegraphics[width=0.70\textwidth]{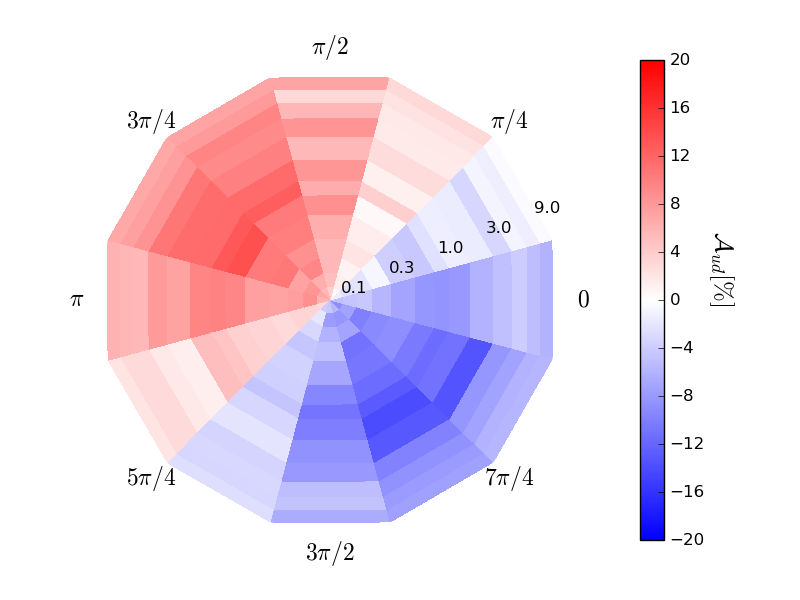}
\caption{
Up-down asymmetry \Aud for simulated samples of $B^+ \rightarrow K_1(1270)^+ \gamma$ decays governed by two amplitudes only, $K_1(1270)^+ \to \Kp\rho(770)^0$ and $K_1(1270)^+  \rightarrow K^{*}(892)^0 \pi^+$, shown as a function of the generated ratio of fractions (radial coordinate, from 0.1 to 9.0) and phase difference between the two amplitudes (polar coordinate).}
\label{fig:k1270_kstpi_k1270_rhok_aud}
\end{figure}

The effects of the resonant structure of the \Kpipi system on $\mathcal{A}_{\text{ud}}$ can be illustrated using a simplified \mbox{\decay{\Bp}{K^+_{\text{res}} \gamma}} model containing only two amplitudes corresponding to the decays \linebreak \mbox{$K_1(1270)^+\!\to\Kp \rho(770)^0\!\to \Kp \pim \pip$} and $K_1(1270)^+\!\to K^{*}(892)^0 \pi^+\!\to \Kp \pim \pip$.
Simulated samples of decays containing only right-handed photons are generated with different relative fractions (as defined in Eq.~\ref{eq:fraction}) and phase differences between these amplitudes, and the up-down asymmetry is computed for each of them.
The results in Fig.~\ref{fig:k1270_kstpi_k1270_rhok_aud} show that the up-down asymmetry varies widely depending on the phase difference between the amplitudes, while it is less dependent on the relative fraction.
This implies that, even in this simple model, the proportionality coefficient that relates the up-down asymmetry to the photon polarisation parameter depends strongly on the phase difference between the amplitudes, making the knowledge of this phase essential to measure the value of $\lambda_{\gamma}$;
additionally, for some values of the relative phase, the proportionality coefficient is null, indicating that the measurement of the up-down asymmetry is not sensitive to $\lambda_{\gamma}$ in such configurations.

To overcome these difficulties and measure the photon polarisation, we propose an analysis that combines information from the angular variables and the squared invariant-mass distributions in order to characterise the interferences between decay processes and their effect on~\lg.

\section{Method\label{sec:method}}

\subsection{Photon polarisation parameter}

The differential decay rate for $\BtoKpipig$ decays that proceed through a single resonance $K_{\text{res}}^{i}$ can be written as \cite{Gronau:2002rz}

\begin{equation}\label{eq:DiffDecaySingleAmp}
\dfrac{d\Gamma (B\rightarrow K_{\text{res}}^{i} (\to K \pi \pi ) \gamma)}{ds_{123}} = |c^{i}_{\textrm R} \mathcal{T}^{i}(s_{123}) A^{i}_{\textrm R}|^{2} + |c^{i}_{\textrm L} \mathcal{T}^{i}(s_{123}) A^{i}_{\textrm L} |^{2}\,,
\end{equation}
where $s_{123}$ is the invariant mass of the $K \pi \pi$ system, $c^{i}_{\textrm R}$ and $c^{i}_{\textrm L}$ are the right- and left-handed weak radiative decay amplitudes, $\mathcal{T}^{i}(s_{123})$ is the propagator associated to resonance $K_{\text{res}}^{i}$, and $A^{i}_{\textrm R}$ and $A^{i}_{\textrm L}$ are the strong decay amplitudes for $K_{\text{res, R/L}}^{i} \rightarrow K \pi \pi$.
The right- and left-handed amplitudes do not interfere since the photon polarisation is an observable quantity. 
For a given resonance $K_{\text{res}}^{i}$, a photon polarisation parameter $\lambda^{i}_{\gamma}$ is defined in terms of the weak radiative decay amplitudes,

\begin{equation}
    \lambda_\gamma^{i}\equiv\frac{|c^{i}_{\textrm R}|^2-|c^{i}_{\textrm L}|^2}{|c^{i}_{\textrm R}|^2+|c^{i}_{\textrm L}|^2}\,.
\end{equation}

Using an argument of parity invariance in strong interactions, detailed in Ref.~\cite{becirevic2012}, the weak radiative decay amplitudes associated with a resonance $K_{\text{res}}^{i}$ in decays of a $B^{+}$ or $B^{0}$ meson can be written as~\cite{Gronau:2002rz,Paul:2016urs}

\begin{equation}\label{eq:crAsFunctionOfC7}
    \begin{pmatrix} c^{i}_{\textrm R} \\ c^{i}_{\textrm L}  \end{pmatrix}  = - \dfrac{4G_{\textrm F}}{\sqrt{2}} V_{tb} V_{ts}^{*} \begin{pmatrix} C_{7}^{\text{eff}}\,g^{i}(0) + h_{\text{R}}^{i}  \\ C'_{7}\,P_{i} (-1)^{J_{i}-1}\,g^{i}(0) + h_{\text{L}}^{i} \end{pmatrix}\,,
\end{equation}
where $G_{\textrm F}$ is the Fermi constant, $V_{tb}$ and $V_{ts}^{*}$ are CKM matrix elements, $P_{i}$ and $J_{i}$ are the parity and spin of the $K_{\text{res}}^{i}$ resonance, $g^{i}(0)$ is the process-dependent hadronic form factor,  $C^{\text{eff}}_{\textrm 7}$ and $C'_{7}$ are the radiative Wilson coefficients, and the quantities $h_{\text{R/L}}^{i}$ encode remaining contributions from the $Q_{1-6}$ and $Q_8$ hadronic operators (see Ref.~\cite{Paul:2016urs} for more details). The coefficient $C^{\text{eff}}_{7}$ includes ``effective'' linear contributions from the other coefficients $C_{1-6}$ in order to make it regularisation- and renormalisation-scheme independent, as discussed in Ref.~\cite{Chetyrkin:1996vx}.
Assuming that the $h_{\text{R/L}}^{i}$ terms are small enough to be neglected in the expressions of $c^{i}_{\textrm R}$ and $c^{i}_{\textrm L}$, the photon polarisation parameter reduces to

\begin{equation}\label{eq:lambdagamma}
    \lambda_\gamma^{i}=\frac{|C_{7}^{\text{eff}}|^2 - |C'_{7}|^2}{|C_{7}^{\text{eff}}|^2 + |C'_{7}|^2}\equiv \lg \,,
\end{equation}

\noindent \ie, the photon polarisation in the weak decay \BptoKresg is the same for all kaonic resonances $K_{\text{res}}^{i}$ and it can be expressed only as a function of Wilson coefficients.\footnote{It is sufficient to assume that the ratio $h_{\text{R/L}}^{i}/g^{i}(0)$ is process independent to enable the definition of a photon polarisation parameter that does not depend on the kaonic resonance $K_{\text{res}}^{i}$.
Actually, differences in $g^{i}(0)$ and $h_{\text{R/L}}^{i}$ between the considered kaonic resonances should be small, as spectator scattering and weak annihilation corrections are expected to be similar amongst the considered resonances, leaving mainly soft gluon corrections to quark loop spectator scattering as the main source of differences.
These latter corrections would need to be taken into account when translating the measurement of the photon polarisation to constraints on the Wilson coefficients.}
In the SM, the value of \lg is expected to be $+1$ (up to corrections of the order of $m_s^2/m_b^2$) for decays of a $B^+$ or $B^{0}$ meson while it is expected to be $-1$ for decays of a $B^-$ or $\bar{B}^{0}$ meson.

\subsection{Amplitude formalism}

To develop our formalism, decays of $B$ mesons to $K\pi\pi\gamma$ are assumed to proceed through a cascade of quasi-independent two-body decays, an approximation known as the \textit{isobar model}~\cite{PhysRev.123.333,PhysRevD.11.3165}. 
In this study, decay topologies of the form $B \rightarrow R_i \gamma$, $R_i \rightarrow R_j P_{1} $, and $R_j \rightarrow P_{2} P_{3}$ are considered, where $R_i$ is a $K\pi\pi$ intermediate state, $R_j$ is either a $K\pi$ or $\pi\pi$ resonant state and $P_{\alpha}$ is a final-state kaon or pion.
The function used to describe \BtoKpipig decays with the above topologies is therefore written as

\begin{equation}\label{eq:signalFunction}
 \mathcal{P}_{\text{s}}  = \dfrac{(1+\lg)}{2}|\mathcal{M}_{\textrm R}|^{2} +  \dfrac{(1-\lg)}{2} |\mathcal{M}_{\textrm L}|^{2}\,,
\end{equation}
where amplitudes for various decay modes associated with right-handed (or left-handed) photons are summed coherently,

\begin{equation}
\mathcal{M}_{\textrm R/\textrm L} = \sum_{k} f_{k}\mathcal{A}_{k, \textrm R/\textrm L}(\bm{x})\quad\quad\text{with}\quad f_{k}=a_{k}e^{i\phi_{k}}\,.
\end{equation}

The decay amplitude $\mathcal{A}_{k, \textrm R/\textrm L}(\bm{x})$ corresponds to a \BtoKpipig process $k$ involving resonances $R_i$ and $R_j$ and a right- or left-handed photon, and $\bm{x}$ is the set of four-vectors associated with the final-state particles in the rest frame of the $B$ meson.
The complex coefficient $f_{k}=a_{k}e^{i\phi_{k}}$ accounts for the magnitude $a_{k}$ and phase $\phi_{k}$ of decay amplitude $k$ and is assumed to be the same for decays with right- or left-handed photons. 
The amplitude for a given decay mode $k$ is a product of resonance propagators $\mathcal{T}$ for each intermediate two-body decay with relative angular momentum $L$, a normalised Blatt-Weisskopf coefficient $B_{L_B}$ for the two-body decay of the $B$ characterised by relative angular momentum $L_B$ and breakup momentum $q_B$, and an overall spin factor $\mathcal{S}_{ij}$ that encodes the dependence of the amplitudes on angular momenta,

\begin{equation}
\mathcal{A}_{\textrm R}^{k}(\bm{x})  = B_{L_B}(q_B(\bm{x}),0) \mathcal{T}_{i}^{k}(\bm{x}) \mathcal{T}_{j}^{k}(\bm{x}) \mathcal{S}_{ij,\textrm R}^{k}(\bm{x})\,,
\end{equation}
and
\begin{equation}
\mathcal{A}_{\textrm L}^{k}(\bm{x})  = P_{i} (-1)^{J_{i}-1} B_{L_B}(q_B(\bm{x}),0) \mathcal{T}_{i}^{k}(\bm{x}) \mathcal{T}_{j}^{k}(\bm{x}) \mathcal{S}_{ij, \textrm L}^{k}(\bm{x})\,.
\end{equation}

Resonances are described by the product of a normalised Blatt-Weisskopf coefficient and a relativistic Breit-Wigner \cite{Jackson:1964zd} lineshape,\footnote{Alternative lineshapes, such as the Gounaris-Sakurai one \cite{Gounaris:1968}, may be more adequate to describe certain resonances, but for simplicity only Breit-Wigner lineshapes are used in the study presented here.} 

\begin{equation}
 \mathcal{T} (s,q,L) =  \dfrac{\sqrt{c}\hspace{5pt}B_{L}(q,0)}{m_0^2 - s - im_{0}\Gamma (s,q,L)}\,,
\end{equation}
where $m_0$ is the nominal mass of the resonance, $q$ denotes the breakup momentum of the outgoing particle pair in the rest frame of the resonance and $\Gamma (s,q,L)$ is its energy-dependent width. 
The normalisation constant
\begin{equation}
 c =  \dfrac{m_{0}\Gamma_{0}\gamma_{0}}{\sqrt{m_{0}^2 + \gamma_{0}}}, \text{ with } \gamma_{0} = m_{0}\sqrt{m_{0}^2 + \Gamma_{0}^2}\,,
\end{equation}

reduces correlations between the coupling to the decay channel and the mass and width of the resonance. 
The width of the resonance for a decay into two particles is parametrised as

\begin{equation}
 \Gamma (s,q,L) = \Gamma_{0} \dfrac{m_0}{\sqrt{s}}\bigg(\dfrac{q}{q_0}\bigg)^{2L+1} B_{L}(q,q_{0})^2\,,
\end{equation}
where $q_0$ is the value of the breakup momentum at the resonance pole $s={m_0}^{2}$, and $B_{L}(q,q_{0})$ is the normalised Blatt-Weisskopf barrier factor, listed in Table~\ref{tab:barrierFactors}.

\begin{table}[tb]
    \centering
    \caption{Normalised Blatt-Weisskopf centrifugal barrier factors for angular momentum $L$. The meson radial parameter $R$ is set to $1.5\,(\text{GeV}/c)^{-1}$ following a measurement by Belle~\cite{PhysRevD.83.032005}.}
    \medskip
    \begin{tabular}{lc}
        \toprule
        $L$ & $B_{L} (q,q_0)$ \\
        \midrule
        $0$ & $1 $ \\
        \addlinespace[0.1cm]
        $1$ & $\sqrt{\dfrac{1+R^2q_{0}^{2}}{1+R^2q^2}} $ \\
        \addlinespace[0.1cm]
        $2$ & $\sqrt{\dfrac{9+3R^2q_{0}^2+R^4q_{0}^4}{9+3R^2q^2+R^4q^4}} $ \\
        \addlinespace[0.1cm]
        \bottomrule
        \label{tab:barrierFactors}
    \end{tabular}%
\end{table}

The spin factors $\mathcal{S}_{ij,\textrm R/\textrm L}$ are constructed using the Rarita-Schwinger (covariant tensor) formalism, following the method described in Ref.~\cite{dArgent:2017gzv}. 
The spin factors used in this study, as well as a brief description of their computation, are given in Appendix~\ref{app:SpinFactors}. 

\subsection{Amplitude fit}

The proposed method to determine the photon polarisation parameter \lg utilises all the degrees of freedom of the system to perform a maximum likelihood fit to the data using a probability density function (PDF) that depends explicitly on \lg.
This amplitude fit allows the direct measurement of \lg, as well as of the relative magnitudes and phases of the different decay-chain amplitudes included in the model.
The PDF is computed using the function $\mathcal{P}_{\text{s}}$ given in Eq.~\ref{eq:signalFunction} as

\begin{equation}\label{eq:likelihood}
\mathcal{F}(\bm{x} | \Omega) = \dfrac{\xi(\bm{x}) \mathcal{P}_{\text{s}}(\bm{x} | \Omega) \Phi_{4}(\bm{x}) }{\int \xi(\bm{x}) \mathcal{P}_{\text{s}}(\bm{x} | \Omega)\Phi_{4}(\bm{x}) \diff \bm{x}}  \,,
\end{equation}
where $\Omega = \lambda_{\gamma}, \{a_{k}\},\{\phi_{k}\} $ is the set of fit parameters, $\Phi_{4}(\bm{x})$ is the four-body phase-space density, and $\xi(\bm{x})$ is the efficiency, which accounts for effects related to detector acceptance, reconstruction, and event selection.

The magnitude and phase of each amplitude $k$ ($a_{k}$ and $\phi_{k}$) are measured with respect to those of amplitude $1$, for which $a_{1}$ and $\phi_{1}$ are fixed to $1$ and $0$, respectively.

The normalisation integral of Eq.~\ref{eq:likelihood} is computed numerically using a large sample of simulated events, generated according to an approximate model $\mathcal{P}_{\text{gen}}$.
The signal acceptance $\xi(\bm{x})$ is inherently taken into account by applying the event selection used in data to these simulated events; the normalisation integral can then be estimated as
\begin{equation}
\int \xi(\bm{x}) \mathcal{P}_{s}(\bm{x}| \Omega) \Phi_{4}(\bm{x}) \diff \bm{x}  = \dfrac{I_{\text{gen}}}{N_{\textrm{sel}}}\sum_{j}^{N_{\textrm{sel}}} \dfrac{\mathcal{P}_{s}(\bm{x}_{j}| \Omega)}{\mathcal{P}_{\text{gen}}(\bm{x}_{j})}\quad\text{with } I_{\text{gen}} = \int  \xi(\bm{x}) \mathcal{P}_{\text{gen}}(\bm{x}) \Phi_{4}(\bm{x}) \diff \bm{x} \,,
\end{equation}
where $N_{\text{sel}}$ is the total number of generated events that pass the selection criteria.
Note that $I_{\text{gen}}$ does not depend on the parameters of the fit, and therefore does not need to be evaluated to perform the maximisation.

For the studies presented here, the effect of the application of a selection is not considered, \ie, $\xi(\bm{x}) = 1$.

The fraction of a decay mode $k$ is defined as the ratio of the phase-space integral of the sum of right- and left-handed contributions over the phase-space integral of the function $\mathcal{P}_{\text{s}}$,

\begin{equation}\label{eq:fraction}
    F_{k} = \frac { \int \left\{(1+\lg ) |f_{k}\mathcal{A}_{k, \textrm R}(\bm{x})|^2 +  (1-\lg) |f_{k}\mathcal{A}_{k, \textrm L}(\bm{x})|^2\right\} \Phi_{4}(\bm{x}) \diff\bm{x}}{2 \int \mathcal{P}_{s}(\bm{x}| \Omega)  \Phi_{4}(\bm{x}) \diff \bm{x} } \,.
\end{equation}
Due to interferences between the decay modes, the sum of these fractions may not be equal to unity.
The interference term between the decay modes $k$ and $l$, where $k > l$, can be expressed as

\begin{equation}
F_{kl} =  \frac{\int \left\{(1+\lg) \mathcal{R}e( f_{k}\mathcal{A}_{k, \textrm R}(\bm{x})f^*_{l}\mathcal{A}^*_{l, \textrm R}(\bm{x})) + (1-\lg) \mathcal{R}e( f_{k}\mathcal{A}_{k, \textrm L}(\bm{x})f^*_{l}\mathcal{A}^*_{l, \textrm L}(\bm{x}))\right\} \Phi_{4}(\bm{x}) \diff\bm{x}}{\int \mathcal{P}_{s}(\bm{x}| \Omega)  \Phi_{4}(\bm{x}) \diff \bm{x} }\,,
\end{equation}
such that the sum of all the fractions and interference terms is equal to unity:
\begin{equation}
\sum_k F_{k} + \sum_{k > l} F_{kl} = 1\,.
\end{equation}

\section{Sensitivity\label{sec:sensitivity}}

The amplitude formalism described in Sec.~\ref{sec:method} is implemented in a generator and fitter software framework developed for the amplitude analysis of $D^0 \rightarrow K^+ K^- \pip\pim$ decays at CLEO \cite{Artuso:2012df,dArgent:2017gzv}. 
The performance of the amplitude fitter is studied initially by generating and subsequently fitting simulated data sets of \text{\BtoKpipig} decays using models containing two or three amplitudes.
Once the methodology is validated, more realistic models of the \Kpipi system are used in order to obtain prospects for measurements of the photon polarisation parameter in $B$-physics experiments.

\subsection{Proof-of-concept using simplified models\label{subsec:proof-of-concept}}

As illustrated in Fig.~\ref{fig:k1270_kstpi_k1270_rhok_aud}, the sensitivity to the photon polarisation parameter obtained from the up-down asymmetry depends primarily on the relative phase.
The same set of simplified models of the \decay{\Bp}{K_1(1270)^+\gamma} channel, which include only two decay modes of the kaonic resonance (\decay{K_1(1270)^+}{\Kp\rho(770)^0} and \decay{K_1(1270)^+}{K^{*}(892)^0\pip}) is used to test the performance of the full amplitude fit, as well as its stability and the accuracy of the obtained uncertainties.
The free parameters of the fit are the photon polarisation parameter \lg, and the modulus and phase associated with the \decay{K_1(1270)^+}{\Kp\rho(770)^0} channel, hereafter referred to as the relative magnitude and phase, where the \decay{K_1(1270)^+}{K^{*}(892)^0\pip} channel is chosen as a reference.

\begin{figure}[tb]
\centering
\includegraphics[width=0.70\textwidth]{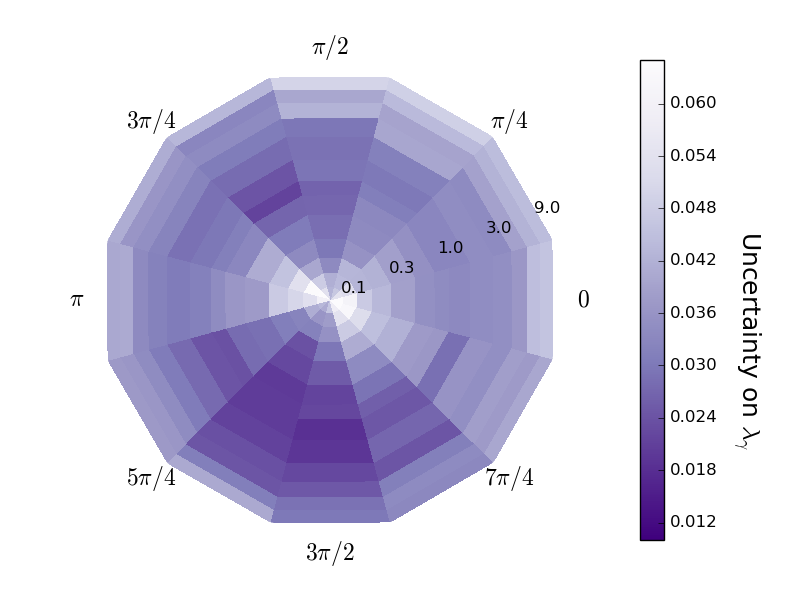}
\caption{%
Uncertainty on \lg obtained from the average of $10$ amplitude fits of simulated samples of $B \rightarrow K_1(1270)^+ \gamma$ decays governed by two amplitudes only, $K_1(1270)^+ \to \Kp\rho(770)^0$ and $K_1(1270)^+  \rightarrow K^{*}(892)^0 \pi^+$, shown as a function of the relative fraction (radial coordinate, from $0.1$ to $9.0$) and phase (polar coordinate) of the two amplitudes.}
\label{fig:k1270_kstpi_k1270_rhok_crsq}
\end{figure}

For several pairs of relative magnitude and phase, $10$ simulated data sets of $\numprint{8000}$ events are generated with $\lg=+1$ (close to the SM value) and fitted independently. 
The average uncertainty on \lg as a function of relative fraction (as defined in Eq.~\ref{eq:fraction}) and phase is shown in Fig.~\ref{fig:k1270_kstpi_k1270_rhok_crsq}, where areas of higher colour saturation indicate regions with higher sensitivity to \lg: unlike \Aud, the amplitude analysis is sensitive to \lg for all values of relative fractions and phases, with statistical uncertainties ranging from $0.01$ to $0.05$.

A higher average uncertainty on \lg is seen for models in which the fraction of one amplitude is much larger than the other, and the maximum sensitivity is obtained for a phase difference of around $3\pi/2$ and a relative fraction of $1.5$.

\begin{figure}[tb]
\centering
\includegraphics[width=0.45\textwidth]{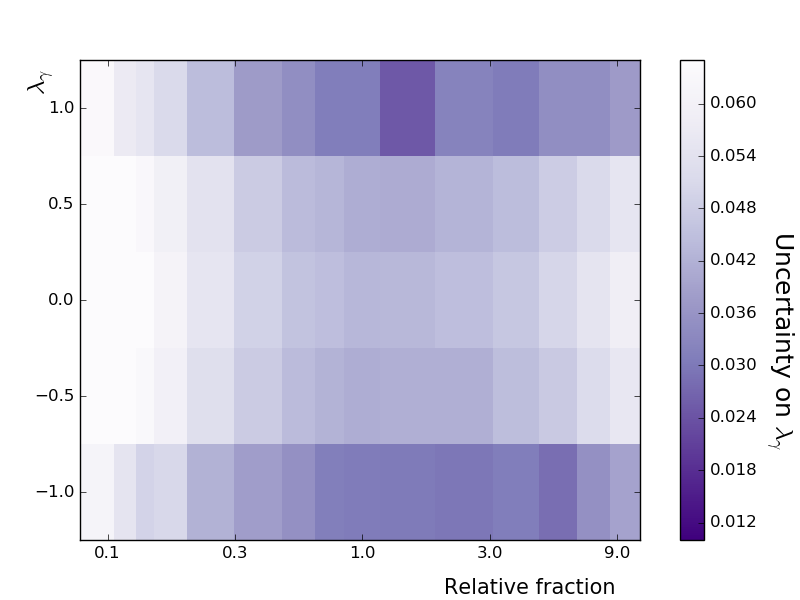}
\includegraphics[width=0.45\textwidth]{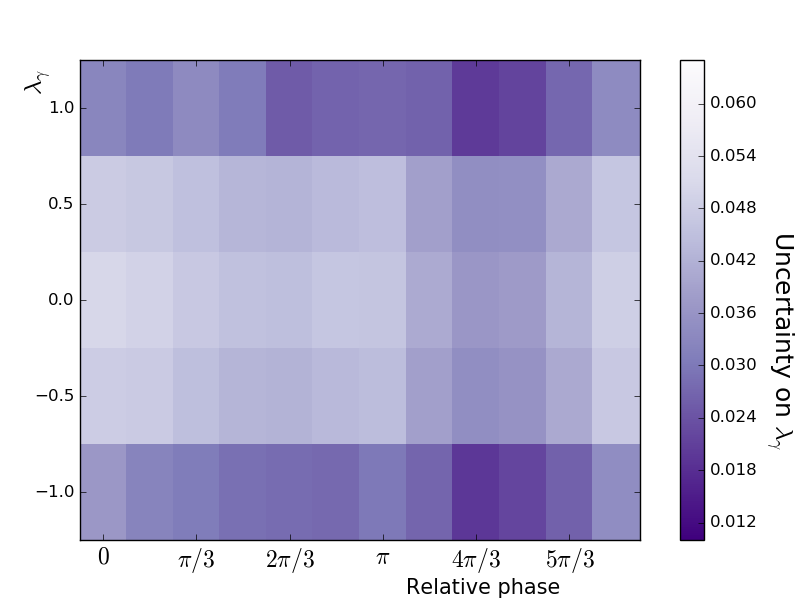}
\caption{
Uncertainty on \lg obtained from the average of 10 amplitude fits of simulated samples of $B \rightarrow K_1(1270)^+ \gamma$ decays governed by two amplitudes only, $K_1(1270)^+ \to \Kp\rho(770)^0$ and $K_1(1270)^+  \rightarrow K^{*}(892)^0 \pi^+$, shown as a function of the generated \lg value and the relative fraction (left) or phase difference (right) of the two amplitudes.}

\label{fig:scan_ampPhase_crsq}
\end{figure}

To evaluate the performance of the fit as a function of the photon polarisation parameter, the study is repeated for various generated values of \lg, and the results are shown in Fig.~\ref{fig:scan_ampPhase_crsq}.
The highest sensitivities to \lg are obtained for $\lg=\pm1$, with increasing uncertainties observed as the generated absolute value of \lg decreases.

To study the fit accuracy and error estimation, $100$ simulated data sets are generated and fitted for selected values of the model parameters (relative magnitude, relative phase and \lg).
As asymmetric errors are used in these fits, the quality of the parameter estimation is evaluated by checking that the distribution of the pull variable $g$ is compatible with a standard normal distribution, where $g$ is defined as:
\begin{align}
\textrm{if (fit result)}\leq\textrm{(true value):}\quad&g=\frac{\textrm{(true value)}-\textrm{(fit result)}}{\textrm{(positive error)}}\,,\\
\textrm{otherwise:}\quad&g=\frac{\textrm{(fit result)}-\textrm{(true value)}}{\textrm{(negative error)}}\,.
\end{align}

The mean values and standard deviations of the fitted parameters and the associated pull parameters can be found in Tables~\ref{tab:fitResults_2amps_RH}, ~\ref{tab:fitResults_2amps_highAud}, and~\ref{tab:fitResults_2amps_lowAud} of Appendix~\ref{app:Sensitivity}.
For all models, each fit parameter has a Gaussian distribution centered on the generated value with a pull distribution of width consistent with unity, resulting in an unbiased measurement and correct error estimation.

As a final test, we study decays of $B$ mesons to $K\pi\pi\gamma$ with a $\pi^0$ in the final state, which can have an additional source of interference from intermediate states that include a $K^{*}(892)$ resonance.
It has been claimed that the presence of these additional interference terms results in a higher maximum possible up-down asymmetry~\cite{Gronau:2002rz}, and thus that the analysis of \BztoKpipizg decays could be potentially more sensitive to the photon polarisation than that of \BptoKpipig decays.

\begin{figure}[tb]
\centering
\includegraphics[width=0.65\textwidth]{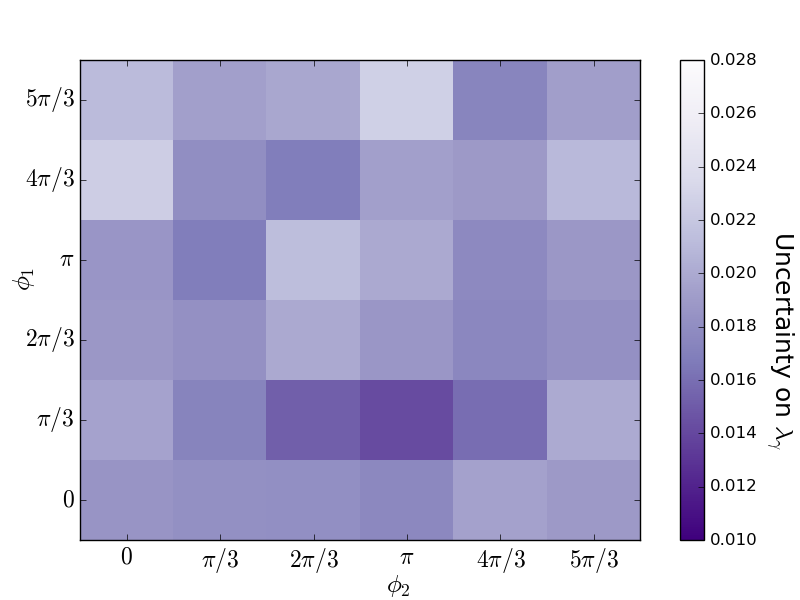}
\caption{
Uncertainty on \lg obtained from the average of $10$ amplitude fits of simulated samples of $B^{0} \rightarrow K_1(1270)^0 \gamma$ decays, shown as a function of the phase differences of the $K_1(1270)^0 \to \Kp\rho(770)^-$ and $K_1(1270)^0  \to K^{*}(892)^0 \pi^0$ decay modes relative to the $K_1(1270)^0  \to K^{*}(892)^+ \pi^-$ decay mode, denoted as $\phi_{1}$ and $\phi_{2}$ respectively.}
\label{fig:neutrals_phaseMap}
\end{figure}

The effect of an additional decay amplitude (and therefore additional interference terms) is studied using a \decay{\Bz}{K_1(1270)^0 \gamma} model with three different $K_1(1270)^0$ decay channels, $K_1(1270)^0 \to \Kp\rho(770)^-$, $K_1(1270)^0  \rightarrow K^{*}(892)^+ \pi^-$, and $K_1(1270)^0  \rightarrow K^{*}(892)^0 \pi^0$.
Ten simulated data sets with $\numprint{8000}$ events each are generated for different values of the phase differences of the $K_1(1270)^0 \to \Kp\rho(770)^-$ and $K_1(1270)^0  \to K^{*}(892)^0 \pi^0$ modes relative to the $K_1(1270)^0  \to K^{*}(892)^+ \pi^-$ mode;
all samples are generated with $\lg=+1$, with the decay rate for all amplitudes being equal. 
The uncertainty on the photon polarisation parameter for all models studied, shown in Fig.~\ref{fig:neutrals_phaseMap}, is within the same range as seen in the two-amplitude $\BptoKpipig$ model, showing that the amplitude analysis is not very sensitive to the number of interference terms in the \Kpipi system.

We conclude that this amplitude analysis is sensitive to the photon polarisation parameter for all simplified models studied, for both charged and neutral decay modes.

\subsection{Prospects for future measurements}

\subsubsection{$\bm{\text{\BptoKpipig}}$ decays}
 
In light of the results of the proof-of-concept model, the most promising measurement of the photon polarisation parameter is expected to come from \BptoKpipig decays, which are the most abundantly reconstructed at \lhcb and \belle II.

\begin{figure}[tbp]
    \centering
    {\includegraphics[width=0.45\linewidth]{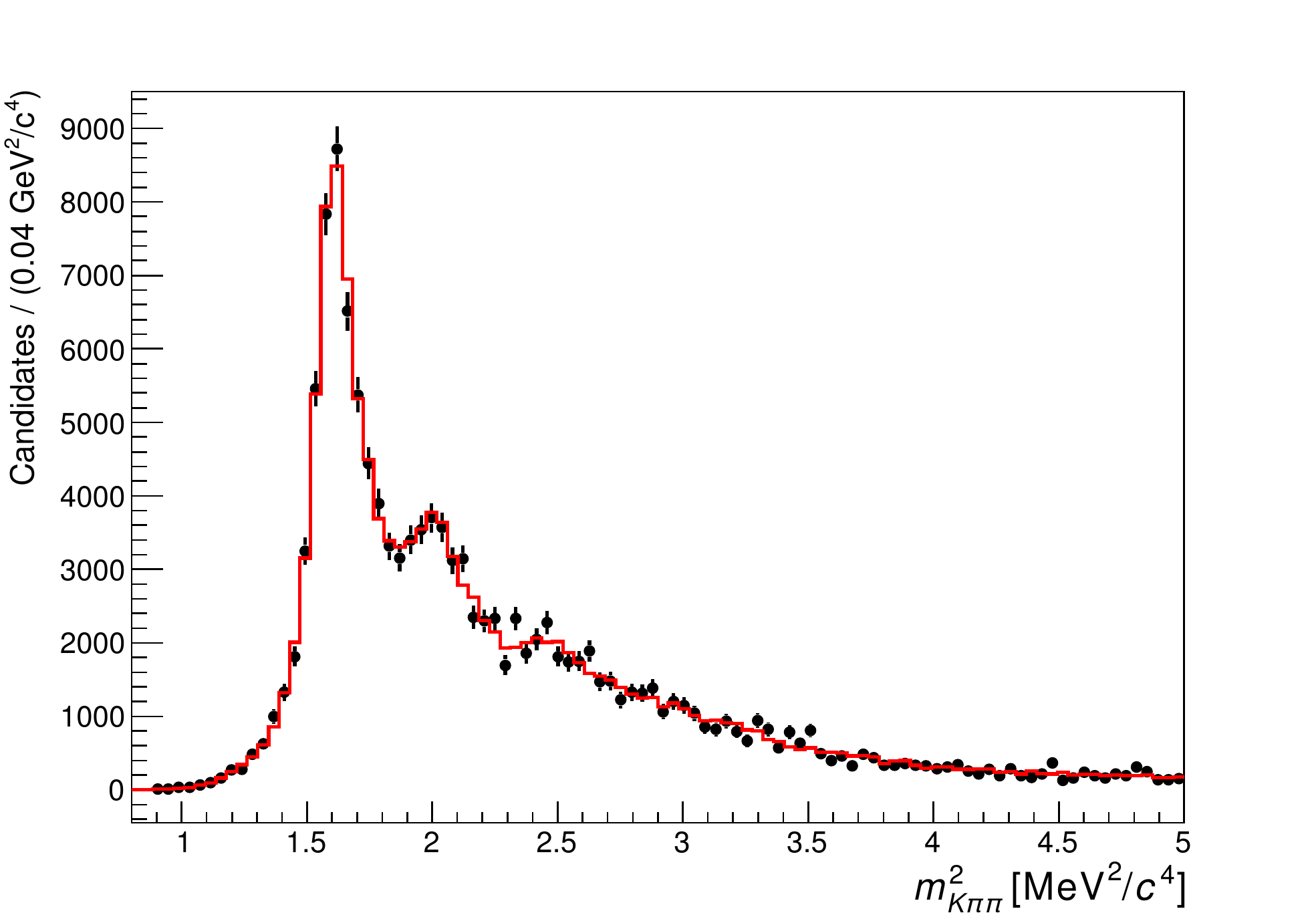}}
    {\includegraphics[width=0.45\linewidth]{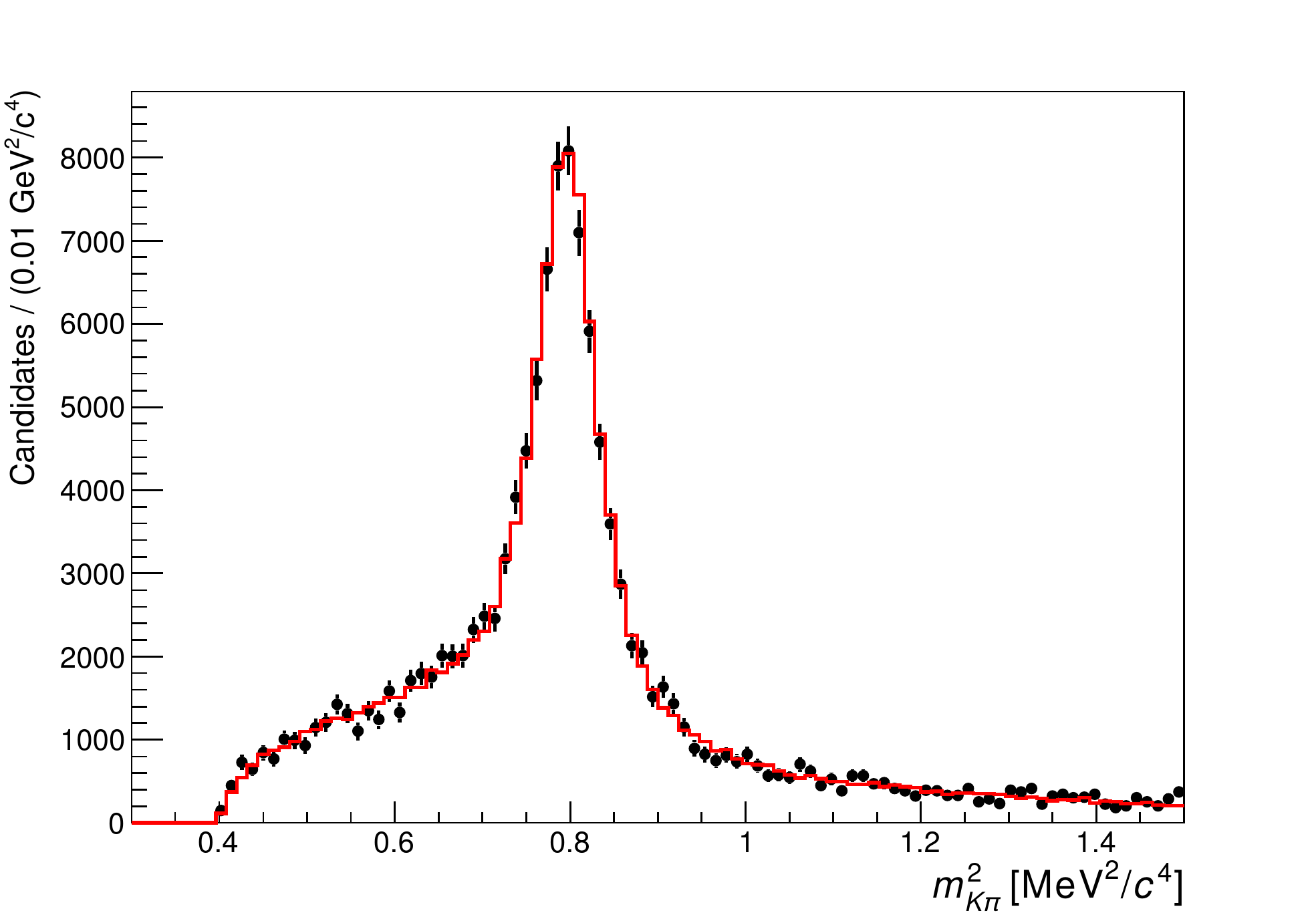}}
    {\includegraphics[width=0.45\linewidth]{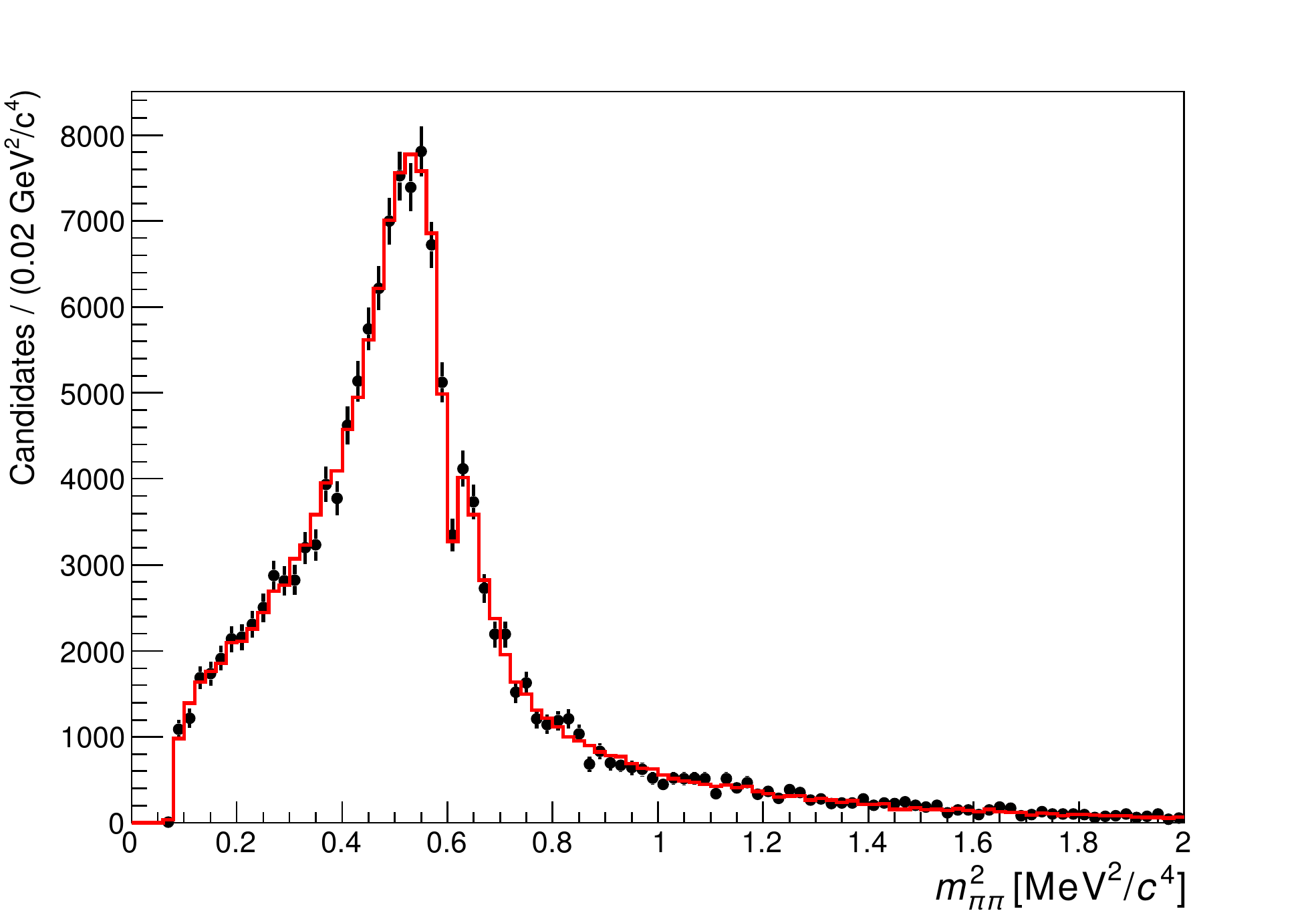}}
    {\includegraphics[width=0.45\linewidth]{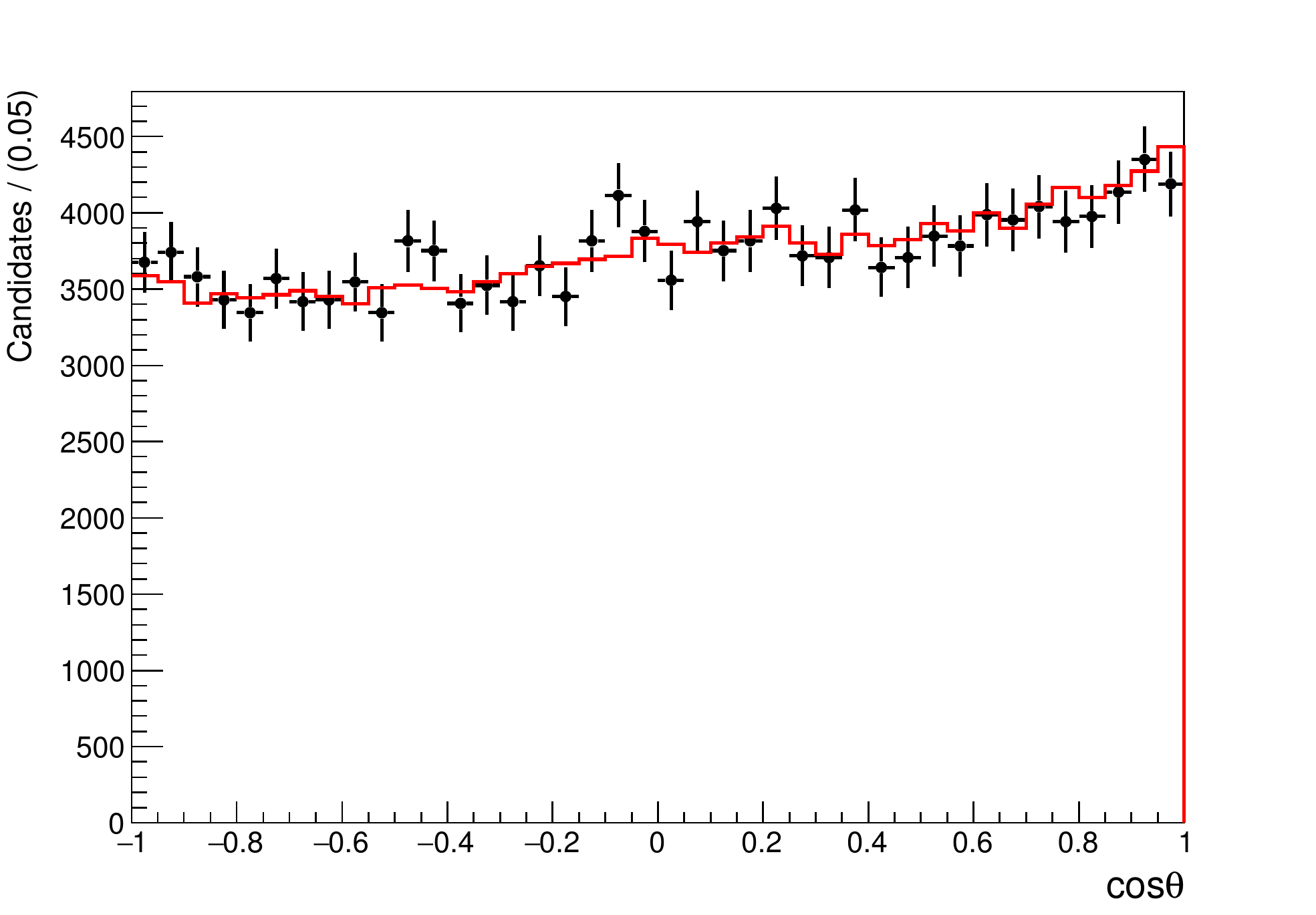}}
    {\includegraphics[width=0.45\linewidth]{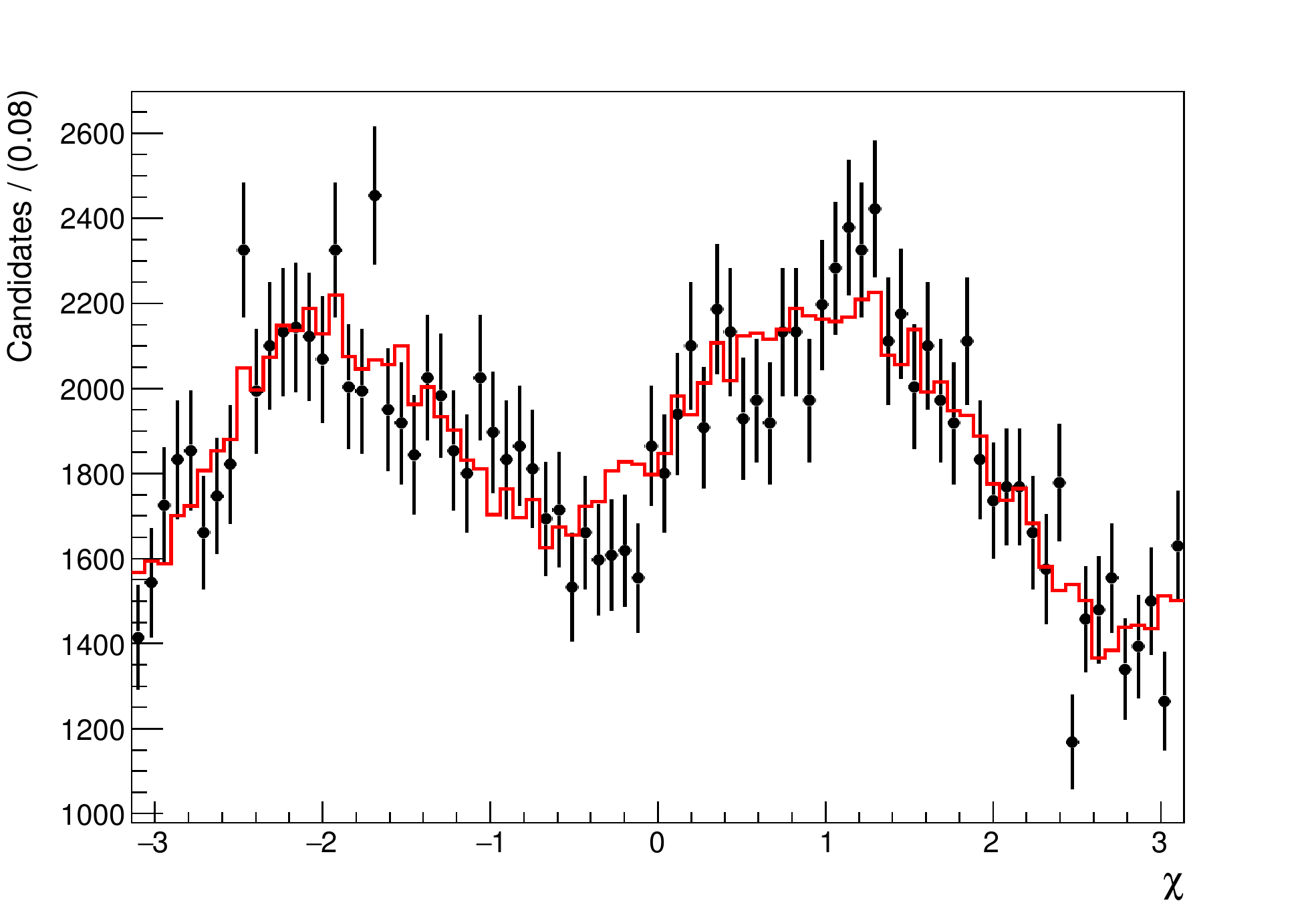}}
    \caption{Squared invariant-mass ($m^2_{K^+\pim\pip}, m^2_{K^+\pim}, m^2_{\pim\pip}$) and angular (cos$\,\theta$ and $\chi$) distributions for a single data set of $\numprint{14000}$ \BptoKpipig decays generated with the $14$ amplitudes listed in Table~\ref{tab:Bp_model_15amps}. The red histograms represent the projections of the PDF obtained from the fit.}
   \label{fig:15ampModel_fitProjections}
\end{figure}

An estimate of the statistical sensitivity of a measurement of the photon polarisation from an amplitude analysis of \BptoKpipig decays is obtained by studying the model described in Table~\ref{tab:Bp_model_15amps}, which provides a good approximation to the $K\pi$, $\pi\pi$ and \Kpipi invariant mass spectra observed in a data sample of $3\,\invfb$ collected by \lhcb during Run 1 of the LHC~\cite{LHCb-PAPER-2014-001,Veneziano:214766}.
A total of $100$ data sets of $\numprint{14000}$ events each, corresponding to the LHCb signal yield of Run 1~\cite{LHCb-PAPER-2014-001}, are generated with $\lg=+1$. The fits of these samples yield a mean uncertainty on \lg of $0.014$.
Figure~\ref{fig:15ampModel_fitProjections} shows the distributions for the five variables for one of these simulated data sets along with the corresponding projections of the fit PDF.
While the pull means ($\mu_{\textrm{pull}}$) and widths ($\sigma_{\textrm{pull}}$) of the complex coefficients $a_{k}$ and $\phi_k$, listed in Table~\ref{tab:fitResults_15amp}, show that the fit is unbiased and the errors are well estimated, the pull distribution associated with \lg has a mean of $0.22 \pm 0.12$ and a width of $1.22 \pm 0.08$, indicating that the obtained uncertainty on \lg is underestimated by about $20\%$.

Taking into account a corrected uncertainty of $0.017$, the comparison of this result with the simplified models discussed in the previous section suggests that the model complexity does not have a large effect on the sensitivity to \lg.\footnote{It is worth noting that more complex models typically entail larger systematic uncertainties, so this conclusion is valid only in what regards the statistical error obtained from the fit.}
This fact can be used to evaluate the gain in sensitivity that could be obtained by exploiting the additional $6\,\invfb$ of data that have been recorded by \lhcb at a $pp$ energy of $13\,$TeV in Run $2$, where the $B$ production cross-section is almost twice that at the Run $1$ energy of $7-8\,$TeV:
assuming that a total of $\numprint{70000}$ signal decays are selected using the \lhcb Run 1 and Run 2 data sets, the resulting corrected statistical uncertainty on the measurement of the photon polarisation parameter could reach~$0.009$.

\begin{table}[tb]
    \caption{Model used to describe the $K_{\text{res}} \rightarrow K^+ \pi^- \pi^+$ hadronic system in the \text{\BptoKpipig} decays. The table is divided in sections according to the spin-parity $J^{P}$ of the $K_{\text{res}}$ resonance. The amplitude with the S-wave decay $K_1(1270)^+ \to K^{*}(892)^0 \pi^+$ is chosen as a reference for the magnitudes and phases.\label{tab:Bp_model_15amps}}
    \begin{center}
    \begin{tabular}{clccc}
        \toprule
        $J^{P}$ & Amplitude $k$      & $a_{k}$     & $\phi_{k}$ & Fraction ($\%$) \\
        \midrule
        \multirow{4}{*}{$1^{+}$} & \decay{K_1(1270)^+}{K^{*}(892)^0 \pi^+} [S-wave] & $1$ (fixed) & $0$ (fixed)       & $15.3$   \\
        & \decay{K_1(1270)^+}{K^{*}(892)^0 \pi^+} [D-wave] & $1.00$      & $-1.74$           & $\phantom{0}0.6$ \\
        &\decay{K_1(1270)^+}{K^+ \rho(770)^0}             & $2.02$      & $-0.91$           & $37.9$ \\
        & \decay{K_1(1400)^+}{K^{*}(892)^0 \pi^+}          & $0.59$      & $-0.76$           & $\phantom{0}7.4$ \\
        \midrule
       \multirow{3}{*}{$1^{-}$} &  \decay{K^*(1410)^+}{K^{*}(892)^0 \pi^+}          & $0.11$      & $\phantom{-}0.00$ & $\phantom{0}7.9$\\
      &  \decay{K^*(1680)^+}{K^{*}(892)^0 \pi^+}          & $0.05$      & $\phantom{-}0.44$ & $\phantom{0}3.4$\\
       & \decay{K^*(1680)^+}{K^+ \rho(770)^0}             & $0.04$      & $\phantom{-}1.40$ & $\phantom{0}2.3$ \\
        \midrule
      \multirow{2}{*}{$2^{+}$} & \decay{K_2^*(1430)^+}{K^{*}(892)^0 \pi^+$        & }0.28$      & $\phantom{-}0.00$ & $\phantom{0}4.5$\\
       & \decay{K_2^*(1430)^+}{K^+ \rho(770)^0}           & $0.47$      & $\phantom{-}1.80$ & $\phantom{0}8.9$\\
        \midrule
       \multirow{5}{*}{$2^{-}$} & \decay{K_2(1580)^+}{K^*(892)^0 \pi^+}            & $0.49$      & $\phantom{-}2.88$ & $\phantom{0}4.2$ \\
       & \decay{K_2(1580)^+}{K^+ \rho(770)^0}             & $0.38$      & $\phantom{-}2.44$ & $\phantom{0}3.2$\\
       & \decay{K_2(1770)^+}{K^*(892)^0 \pi^+}            & $0.35$      & $\phantom{-}0.00$ & $\phantom{0}2.8$\\
       & \decay{K_2(1770)^+}{K^+ \rho(770)^0}             & $0.08$      & $\phantom{-}2.53$ & $\phantom{0}0.2$\\
       & \decay{K_2(1770)^+}{K_2^*(1430)^0 \pi^+}         & $0.07$      & $-2.06$           & $\phantom{0}0.6$\\
        \bottomrule
    \end{tabular}
    \end{center}
\end{table}

\begin{table}[tb]
    \caption{
    Pull parameters of the fit to \text{\BptoKpipig} samples for all magnitudes and phases relative to the amplitude with the S-wave decay $K_1(1270)^+ \to K^{*}(892)^0 \pi^+$ is chosen as a reference for the magnitudes and phases.\label{tab:fitResults_15amp}}
    \begin{center}
    \begin{tabular}{lcccc}
        \toprule
        Amplitude $k$                                      & \multicolumn{2}{c}{Magnitude $a_{k}$}  & \multicolumn{2}{c}{Phase $\phi_{k}$}  \\
                                                           & $\mu_{\textrm{pull}}$        & $\sigma_{\textrm{pull}}$           & $\mu_{\textrm{pull}}$        & $\sigma_{\textrm{pull}}$ \\ \midrule
        \decay{K_1(1270)^+}{K^{*}(892)^0 \pi^+} [D-wave]   & $\phantom{-}0.12 \pm 0.10 $  & $0.97 \pm 0.07$                    & $-0.01 \pm 0.10 $            & $1.02 \pm 0.07$  \\
        \decay{K_1(1270)^+}{K^+ \rho(770)^0}               & $\phantom{-}0.08 \pm 0.09$   & $0.91 \pm 0.06$                    & $\phantom{-}0.02 \pm 0.11$   & $ 1.08 \pm 0.07 $  \\
        \decay{K_1(1400)^+}{K^{*}(892)^0 \pi^+}            & $-0.44 \pm 0.09 $ & $0.95 \pm 0.06$                    & $\phantom{-}0.87 \pm  0.10$  & $ 1.06 \pm 0.07 $ \\ \midrule
        \decay{K^*(1410)^+}{K^{*}(892)^0 \pi^+}            & $-0.45 \pm 0.09 $            & $0.94 \pm 0.06$                    & $\phantom{-}0.06 \pm 0.10 $  & $  1.04 \pm 0.07 $  \\
        \decay{K^*(1680)^+}{K^{*}(892)^0 \pi^+}            & $\phantom{-}0.04 \pm 0.09 $  & $0.94 \pm  0.06$                   & $\phantom{-}0.02 \pm 0.10 $  & $ 1.08  \pm  0.07$ \\
        \decay{K^*(1680)^+}{K^+ \rho(770)^0}               & $-0.02 \pm 0.11 $            & $1.11 \pm 0.07$                    & $\phantom{-}0.02 \pm 0.10 $  & $ 1.05  \pm 0.07$  \\ \midrule
        \decay{K_2^*(1430)^+}{K^{*}(892)^0 \pi^+}          & $\phantom{-}0.51 \pm 0.10 $  & $1.07 \pm 0.07$                    & $\phantom{-}0.45 \pm 0.09 $  & $ 0.86 \pm 0.06 $  \\
        \decay{K_2^*(1430)^+}{K^+ \rho(770)^0}             & $\phantom{-}0.36 \pm 0.09 $  & $0.98 \pm 0.07$                    & $  -0.01 \pm 0.09 $          & $ 0.94 \pm 0.06$  \\ \midrule
        \decay{K_2(1580)^+}{K^*(892)^0 \pi^+}              & $-0.39 \pm 0.10 $            & $1.03 \pm 0.07$                    & $  -0.06\pm 0.11$            & $ 1.10 \pm 0.07$  \\
        \decay{K_2(1580)^+}{K^+ \rho(770)^0}               & $\phantom{-}0.04 \pm 0.09 $  & $0.90 \pm 0.06$                    & $\phantom{-}0.14 \pm 0.10$   & $  0.97\pm 0.07$  \\
        \decay{K_2(1770)^+}{K^*(892)^0 \pi^+}              & $\phantom{-}0.08 \pm0.11 $   & $1.11 \pm 0.07$                    & $ -0.10 \pm 0.12 $           & $ 1.21 \pm 0.08$  \\
        \decay{K_2(1770)^+}{K^+ \rho(770)^0}               & $-0.13 \pm 0.10$             & $0.97\pm 0.06$                     & $ -0.04 \pm 0.09 $           & $ 0.97  \pm 0.06$  \\
        \decay{K_2(1770)^+}{K_2^*(1430)^0 \pi^+}           & $\phantom{-}0.17 \pm 0.10 $  & $1.05  \pm 0.07$                   & $\phantom{-}0.05  \pm 0.10 $ & $ 1.01 \pm 0.07$  \\
        \bottomrule
    \end{tabular}
    \end{center}
    
\end{table}

\subsubsection{$\bm{\text{\BztoKpipizg}}$ decays}

As discussed in Sec.~\ref{subsec:proof-of-concept}, \text{\BztoKpipizg} decays can also be used for a measurement of the photon polarisation parameter.
The main difference with the \text{\BptoKpipig} decays used above is that the hadronic part of the decays is a priori more complex due to an additional source of interference involving $K^{*}(892)^0 \pi^0$ and $K^{*}(892)^+ \pi^-$ intermediate states in the decays of the heavy kaonic resonances $K_{\text{res}} \to K^+ \pi^- \pi^0$.

\begin{figure}[h]
    \centering
    {\includegraphics[width=0.45\linewidth]{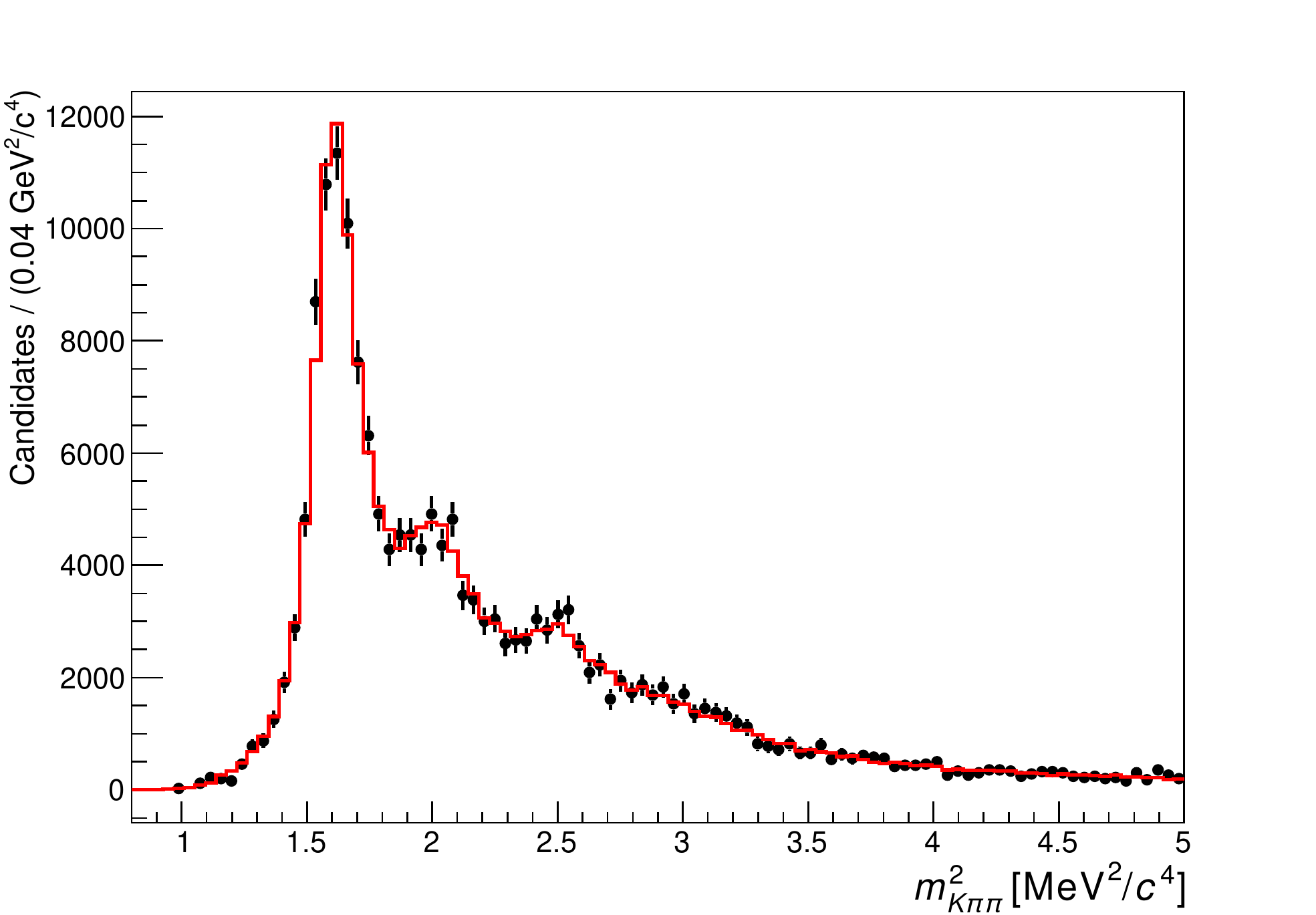}}
    {\includegraphics[width=0.45\linewidth]{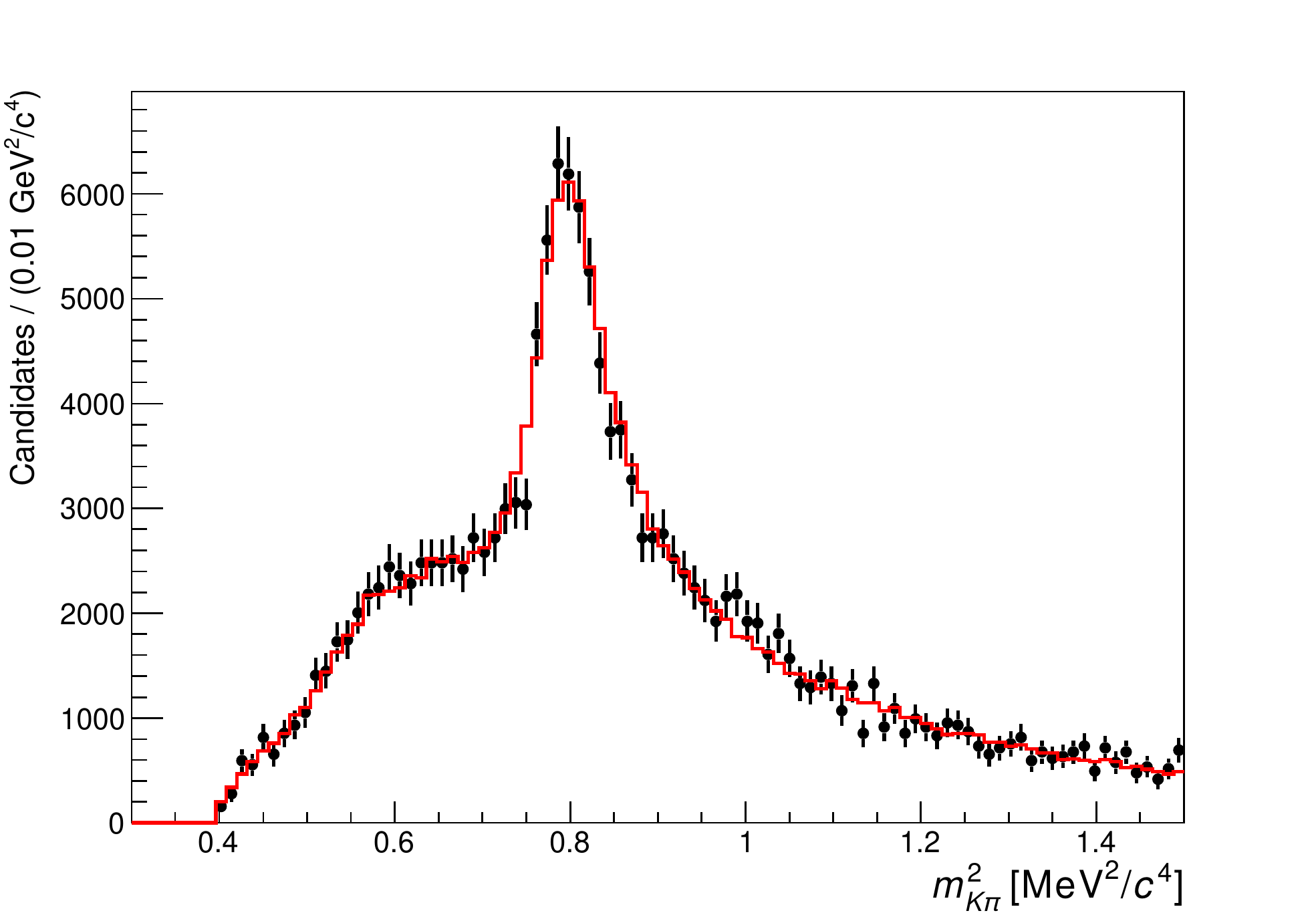}}
    {\includegraphics[width=0.45\linewidth]{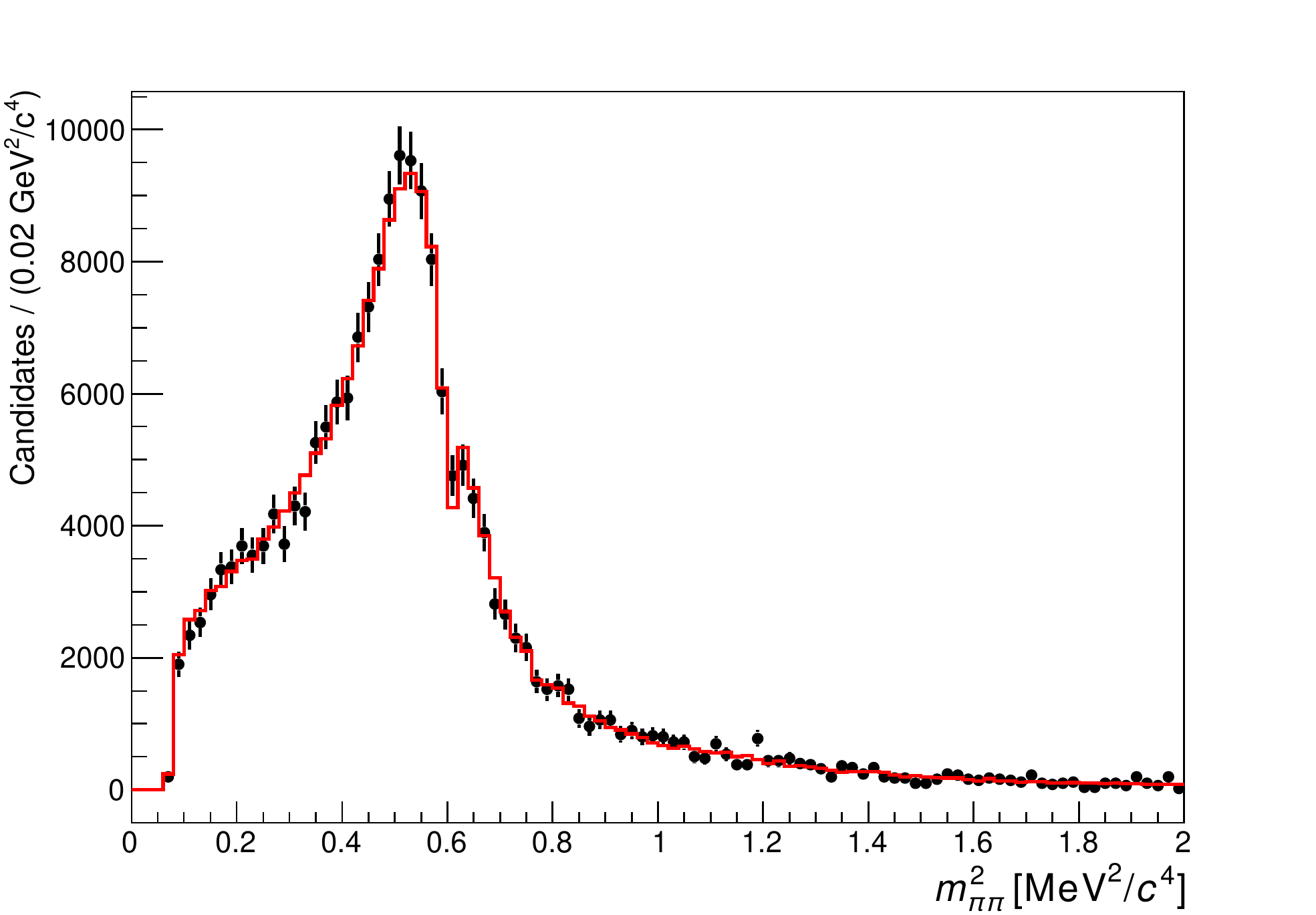}}
    {\includegraphics[width=0.45\linewidth]{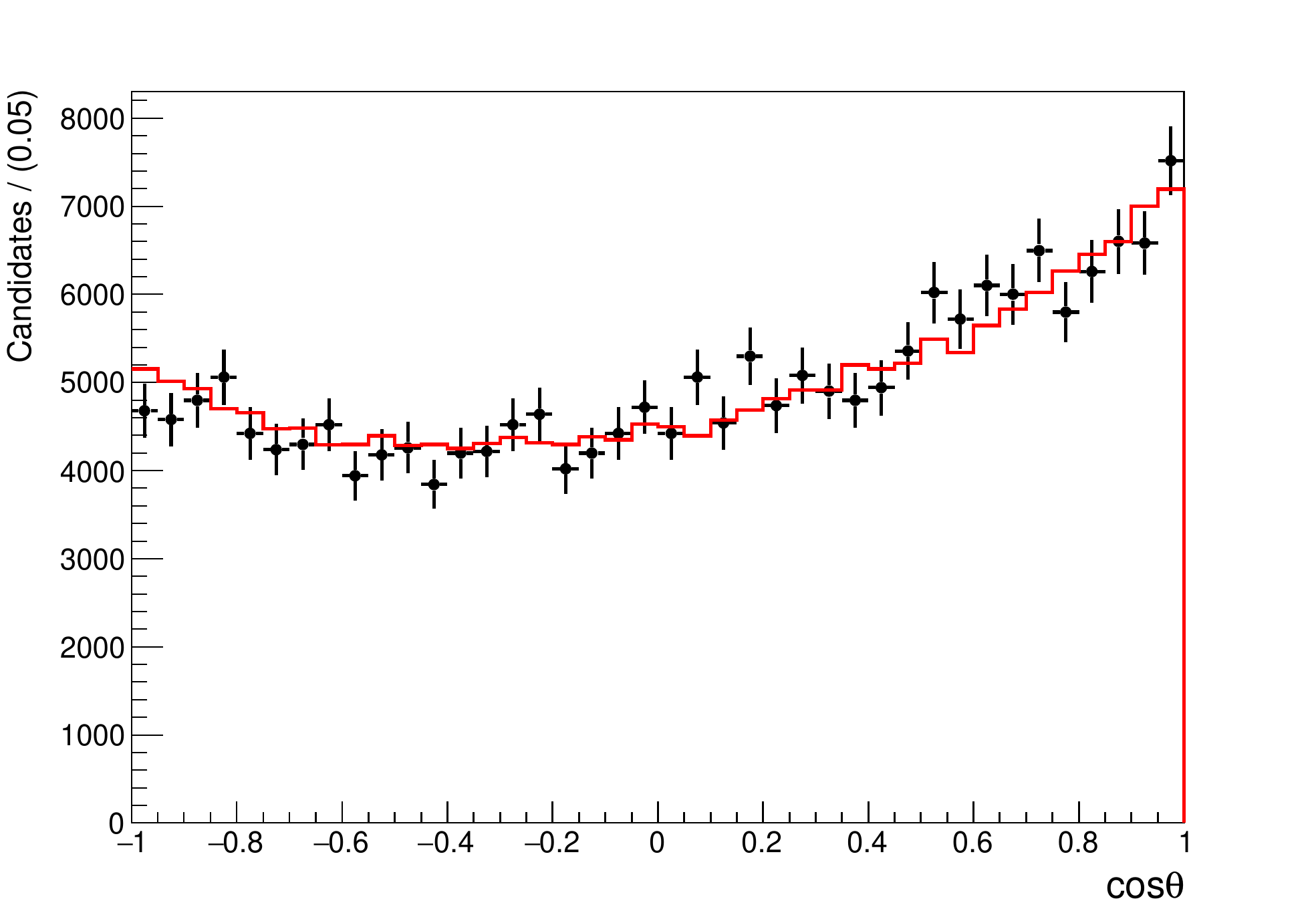}}
    {\includegraphics[width=0.45\linewidth]{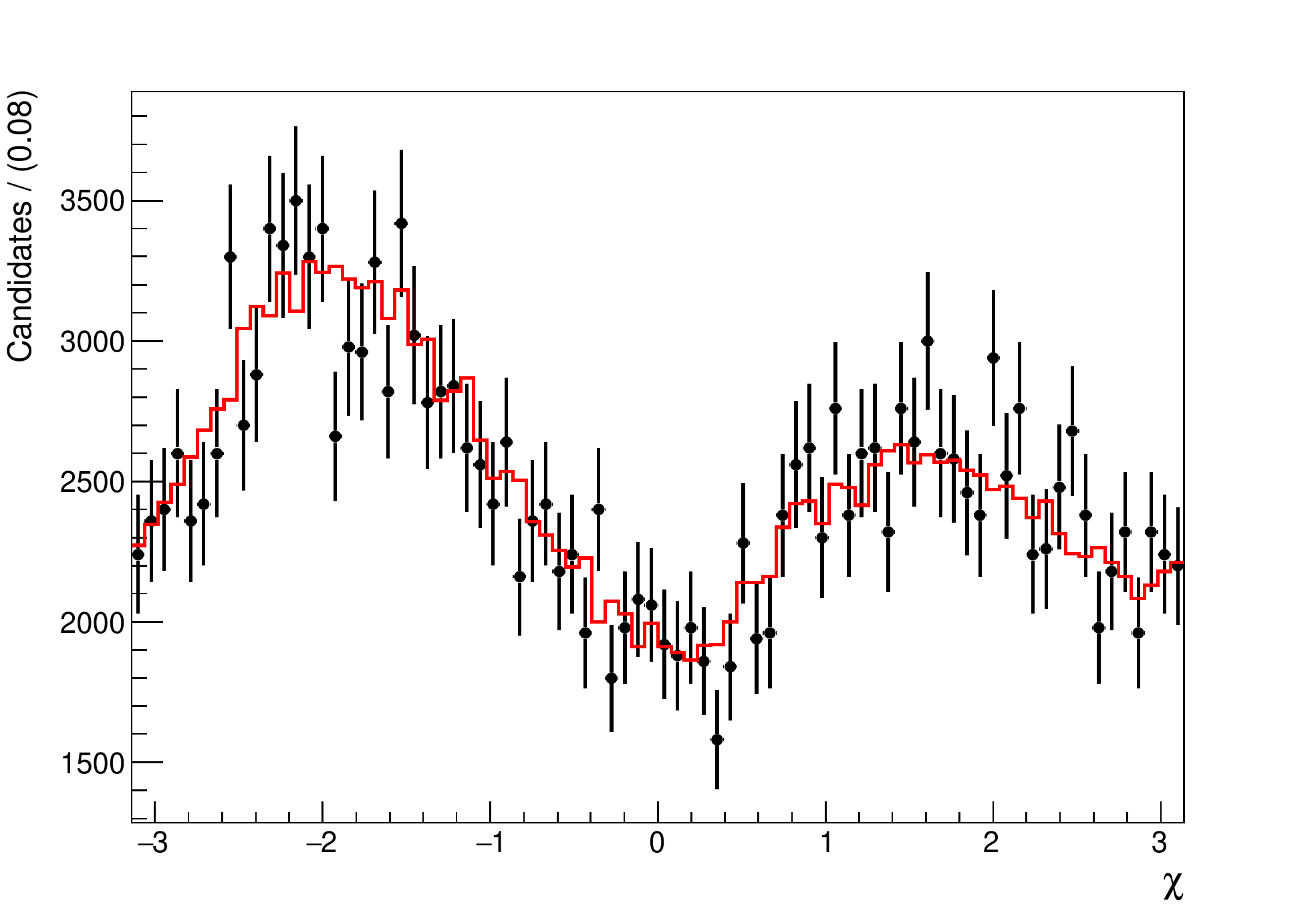}}
    \caption{Squared invariant-mass ($m^2_{K^+\piz\pim}, m^2_{K^+\piz}, m^2_{\piz\pim}$) and angular (cos$\,\theta$ and $\chi$) distributions for a single data set of $\numprint{10000}$ \text{\BztoKpipizg} decays generated with the $23$ amplitudes listed in Table~\ref{tab:model_neutrals}. The red histograms represent the projections of the PDF obtained from the fit.}
   \label{fig:neutralModel_fitProjections}
\end{figure}

In order to evaluate the sensitivity of a measurement of the photon polarisation parameter using \text{\BztoKpipizg} decays, samples of $\numprint{10000}$ simulated signal events (corresponding to the number of expected \text{\BztoKpipizg} decays to be reconstructed by \belle II with $5\,\invab$ of integrated luminosity) are used.
As little is known about the hadronic system in such decays, a model of the \Kpipi system is obtained from the model used for the charged modes, assuming the relative magnitudes and phases of all allowed decay modes without a $K^{*}(892)\pi$ to be identical to those of the charged mode.
In the case of modes with intermediate states that include a kaonic resonance and a pion, the branching fraction is divided equally between the \decay{K_1(1270)^0}{K^{*}(892)^0 (\to\Kp\pim)\piz} and  \decay{K_1(1270)^0}{K^{*}(892)^+(\to\Kp\piz) \pim} modes assuming isospin conservation.
The unknown phase differences are set to the same values for both modes, which is satisfactory in the absence of a strong dependence of the sensitivity of the measurement on the phase difference.
The resulting model, containing 23 amplitudes, is presented in Table~\ref{tab:model_neutrals} and distributions from a single simulated data set are shown in Fig.~\ref{fig:neutralModel_fitProjections}, along with the corresponding fit PDF projections.

Using the same procedure as for the charged mode, an uncertainty on the measurement of the photon polarisation of $0.015$ is obtained from simulated signal samples.
The associated pull width of $1.22 \pm 0.08$ indicates that this uncertainty is also underestimated by around $20\%$; the corrected value of $0.018$ is comparable to the one obtained with the charged mode, confirming that the additional interference patterns and the higher complexity of the \Kpipi system do not provide a significant improvement on the precision of the measurement.
As a higher number of signal events is expected for the charged mode, our method would perform better using these decays, but the amplitude analysis of the neutral mode would provide a very interesting independent measurement of the \lg parameter.

\begin{table}[tb]
    \caption{Model used to describe the $K_{\text{res}} \rightarrow K^+ \pi^- \pi^0$ hadronic system in \text{\BztoKpipizg} decays. The table is divided in sections according to the spin-parity $J^{P}$ of the $K_{\text{res}}$ resonance. The amplitude with the S-wave decay $K_1(1270)^0 \to K^{*}(892)^0 \pi^0$ is chosen as a reference for the magnitudes and phases.}
    \begin{center}
    \begin{tabular}{clccc}
        \toprule
        $J^{P}$ & Amplitude $k$   & $a_{k}$  & $\phi_{k}$ & Fraction ($\%$)\\
        \midrule
       	\multirow{7}{*}{$1^{+}$} & \decay{K_1(1270)^0}{K^{*}(892)^0 \pi^0 } [S-wave] & $1$(fixed) & $0$ (fixed) & $\phantom{0}8.0$\\
        & \decay{K_1(1270)^0}{K^{*}(892)^+ \pi^- } [S-wave] & $1.01$     & $\phantom{-}0.00$     & $\phantom{0}8.0$ \\
        & \decay{K_1(1270)^0}{K^{*}(892)^+ \pi^- } [D-wave] & $0.98$     & $-1.74$     & $\phantom{0}0.3$\\
        & \decay{K_1(1270)^0}{K^{*}(892)^0 \pi^0 } [D-wave] & $0.99$     & $-1.74$     & $\phantom{0}0.3$\\
        & \decay{K_1(1270)^0}{K^+ \rho(770)^- }             & $2.86$     & $-0.91$     & $39.7$\\
        & \decay{K_1(1400)^0}{K^{*}(892)^+ \pi^- }          & $0.60$     & $-0.76$     & $\phantom{0}3.8$ \\
        &\decay{K_1(1400)^0}{K^{*}(892)^0 \pi^0 }          & $0.59$     & $-0.76$     & $\phantom{0}3.8$\\ \midrule
       \multirow{5}{*}{$1^{-}$} & \decay{K^*(1410)^0}{K^{*}(892)^+ \pi^- }          & $0.11$     & $\phantom{-}0.00$      & $\phantom{0}3.9$ \\
       & \decay{K^*(1410)^0}{K^{*}(892)^0 \pi^0 }          & $0.11$     & $\phantom{-}0.00$      & $\phantom{0}3.9$ \\
        &\decay{K^*(1680)^0}{K^{*}(892)^+ \pi^- }          & $0.05$     & $\phantom{-}0.44$      & $\phantom{0}1.7$ \\
       & \decay{K^*(1680)^0}{K^{*}(892)^0 \pi^0 }          & $0.05$     & $\phantom{-}0.44$      & $\phantom{0}1.7$\\
        & \decay{K^*(1680)^0}{K^+ \rho(770)^- }             & $0.06$     & $\phantom{-}1.40$      & $\phantom{0}2.4$\\ \midrule
        \multirow{3}{*}{$2^{+}$} & \decay{K_2^*(1430)^0}{K^{*}(892)^+ \pi^-}         & $0.27$     & $\phantom{-}0.00$      & $\phantom{0}2.3$ \\
       & \decay{K_2^*(1430)^0}{K^{*}(892)^0 \pi^0}         & $0.27$     & $\phantom{-}0.00$      & $\phantom{0}2.3$ \\
       & \decay{K_2^*(1430)^0}{ K^+ \rho(770)^- }          & $0.63$     & $\phantom{-}1.80$      & $\phantom{0}8.9$ \\ \midrule
      \multirow{8}{*}{$2^{-}$} & \decay{K_2(1580)^0}{K^*(892)^+ \pi^- }            & $0.49$     & $\phantom{-}2.88$      & $\phantom{0}2.2$ \\
       & \decay{K_2(1580)^0}{K^*(892)^0 \pi^0 }            & $0.49$     & $\phantom{-}2.88$      & $\phantom{0}2.2$ \\
       & \decay{K_2(1580)^0}{K^+ \rho(770)^- }             & $0.54$     & $\phantom{-}2.44$      & $\phantom{0}3.2$ \\
       & \decay{K_2(1770)^0}{K^*(892)^+ \pi^- }            & $0.35$     & $\phantom{-}0.00$      & $\phantom{0}1.5$ \\
       & \decay{K_2(1770)^0}{K^*(892)^0 \pi^0 }            & $0.35$     & $\phantom{-}0.00$      & $\phantom{0}1.5$\\
       & \decay{K_2(1770)^0}{K^+ \rho(770)^- }             & $0.11$     & $\phantom{-}2.53$      & $\phantom{0}0.2$\\
       & \decay{K_2(1770)^0}{K_2^*(1430)^+ \pi^-}          & $0.07$     & $-2.06$     & $\phantom{0}0.3$\\
       & \decay{K_2(1770)^0}{K_2^*(1430)^0 \pi^0}          & $0.07$     & $-2.06$     & $\phantom{0}0.3$\\
        \bottomrule
    \end{tabular}
    \end{center}
    \label{tab:model_neutrals}
\end{table}

\begin{table}[tb]

    \caption{Pull parameters of the fit to \text{\BztoKpipizg} samples for all magnitudes and phases relative to the amplitude with the S-wave decay $K_1(1270)^0 \to K^{*}(892)^0 \pi^0$.\label{tab:fit_results_neutrals}}
    \begin{center}
     \begin{tabular}{lcccc}
        \toprule
        Amplitude $k$  & \multicolumn{2}{c}{Magnitude $a_{k}$}   & \multicolumn{2}{c}{Phase $\phi_{k}$}  \\
        & $\phantom{-}\mu_{\textrm{pull}}$ & $\sigma_{\textrm{pull}}$ & $\phantom{-}\mu_{\textrm{pull}}$ & $\sigma_{\textrm{pull}}$ \\          
        \midrule
        \decay{K_1(1270)^0}{K^{*}(892)^+ \pi^- } [S-wave] & $\phantom{-}0.08 \pm 0.09$ & $0.95 \pm 0.06$ & $-0.01 \pm 0.08$           & $0.89 \pm 0.06$\\
        \decay{K_1(1270)^0}{K^{*}(892)^+ \pi^- } [D-wave] & $\phantom{-}0.25 \pm 0.08$ & $0.89 \pm 0.06$ & $-0.06 \pm 0.10$           & $1.02 \pm 0.07$\\
        \decay{K_1(1270)^0}{K^{*}(892)^0 \pi^0 } [D-wave] & $-0.14 \pm 0.09$           & $0.97 \pm 0.06$ & $-0.21 \pm 0.11$           & $1.08 \pm 0.07$ \\
        \decay{K_1(1270)^0}{K^+ \rho(770)^0 }             & $\phantom{-}0.14 \pm 0.09$ & $0.95 \pm 0.06$ & $-0.25 \pm 0.09$           & $0.92 \pm 0.06$ \\
        \decay{K_1(1400)^0}{K^{*}(892)^+ \pi^- }          & $\phantom{-}0.13 \pm 0.09$ & $0.90 \pm 0.06$ & $-0.22 \pm 0.09$           & $0.99 \pm 0.06$ \\
        \decay{K_1(1400)^0}{K^{*}(892)^0 \pi^0 }          & $\phantom{-}0.06 \pm 0.10$ & $1.07 \pm 0.07$ & $-0.09 \pm 0.09$           & $0.93 \pm 0.06$ \\ \midrule
        \decay{K^*(1410)^0}{K^{*}(892)^+ \pi^- }          & $\phantom{-}0.16 \pm 0.09$ & $0.93 \pm 0.06$ & $\phantom{-}0.11 \pm 0.09$ & $0.91 \pm 0.06$ \\
        \decay{K^*(1410)^0}{K^{*}(892)^0 \pi^0 }          & $\phantom{-}0.09 \pm 0.10$ & $0.99 \pm 0.07$ & $-0.08 \pm 0.09$           & $0.97 \pm 0.06$ \\
        \decay{K^*(1680)^0}{K^{*}(892)^+ \pi^- }          & $-0.03 \pm 0.10$           & $0.99 \pm 0.06$ & $-0.31 \pm 0.09$           & $0.10 \pm 0.06$ \\
        \decay{K^*(1680)^0}{K^{*}(892)^0 \pi^0 }          & $-0.06 \pm 0.10$           & $1.01 \pm 0.07$ & $\phantom{-}0.07 \pm 0.10$ & $1.03 \pm 0.07$ \\
        \decay{K^*(1680)^0}{K^+ \rho(770)^- }             & $\phantom{-}0.11 \pm 0.10$ & $1.07 \pm 0.07$ & $-0.03 \pm 0.10$           & $1.05 \pm 0.07$ \\ \midrule
        \decay{K_2^*(1430)^0}{K^{*}(892)^+ \pi^-}         & $\phantom{-}0.08 \pm 0.11$ & $1.11\pm 0.07$  & $-0.14 \pm 0.09$           & $0.98 \pm 0.06$ \\
        \decay{K_2^*(1430)^0}{K^{*}(892)^0 \pi^0}         & $-0.12 \pm 0.10$           & $1.07 \pm 0.07$ & $-0.08 \pm 0.10$           & $0.98 \pm 0.06$ \\
        \decay{K_2^*(1430)^0}{ K^+ \rho(770)^- }          & $-0.02 \pm 0.09$           & $0.95 \pm 0.06$ & $-0.11 \pm 0.09$           & $0.98 \pm 0.06$ \\ \midrule
        \decay{K_2(1580)^0}{K^*(892)^+ \pi^- }            & $\phantom{-}0.07 \pm 0.10$ & $1.06 \pm 0.07$ & $-0.14 \pm 0.09$           & $0.92 \pm 0.06$ \\
        \decay{K_2(1580)^0}{K^*(892)^0 \pi^0 }            & $\phantom{-}0.01 \pm 0.09$ & $0.92 \pm 0.06$ & $-0.19 \pm 0.09$           & $0.94 \pm 0.06$ \\
        \decay{K_2(1580)^0}{K^+ \rho(770)^- }             & $-0.18 \pm 0.10$           & $1.06 \pm 0.07$ & $-0.21 \pm 0.10$           & $1.06 \pm 0.07$ \\
        \decay{K_2(1770)^0}{K^*(892)^+ \pi^- }            & $\phantom{-}0.14 \pm 0.09$ & $0.92 \pm 0.06$ & $-0.22 \pm 0.09$           & $0.91 \pm 0.06$ \\
        \decay{K_2(1770)^0}{K^*(892)^0 \pi^0 }            & $\phantom{-}0.10 \pm 0.09$ & $0.96 \pm 0.06$ & $-0.11 \pm 0.09$           & $0.97 \pm 0.06$ \\
        \decay{K_2(1770)^0}{K^+ \rho(770)^- }             & $-0.12 \pm 0.09$           & $0.99 \pm 0.06$ & $\phantom{-}0.15 \pm 0.09$ & $0.92 \pm 0.06$ \\
        \decay{K_2(1770)^0}{K_2^*(1430)^+ \pi^-}          & $\phantom{-}0.15 \pm 0.10$ & $1.03 \pm 0.07$ & $\phantom{-}0.04 \pm 0.08$ & $0.88 \pm 0.06$ \\
        \decay{K_2(1770)^0}{K_2^*(1430)^0 \pi^0}          & $\phantom{-}0.12 \pm 0.10$ & $1.06 \pm 0.07$ & $-0.08 \pm 0.09$           & $0.98 \pm 0.06$ \\
        \bottomrule
    \end{tabular}
    \end{center}
    
\end{table}

\section{Conclusions\label{sec:conclusions}}

A new method to measure the photon polarisation parameter in \BtoKpipig decays from an amplitude analysis is presented.
Using simplified models of the hadronic part of the decay, it is shown that the sensitivity of the photon polarisation parameter measurement does not depend strongly on the configuration or complexity of the \Kpipi system.

The performed studies demonstrate that, in the ideal case of a background-free sample without distortions due to experimental effects, and ignoring the differences between non-factorisable hadronic parameters between the resonances in the \Kpipi system, this method allows the measurement of the photon polarisation with a statistical uncertainty of around $0.009$ on a sample of $\numprint{70000}$ \BptoKpipig decays corresponding to the signal statistics assumed for \lhcb in Runs $1$ and $2$.
Belle II is assumed to reconstruct about $\numprint{10000}$ \BztoKpipizg decays with a data set corresponding to an integrated luminosity of $5\,\invab$.
The analysis of these data could also determine independently the photon polarisation with a statistical uncertainty of the order of 0.018, again ignoring background and experimental effects, as well as non factorisable hadronic uncertainties.

The uncertainty on the measurement of the photon polarisation parameter \lg can be translated in terms of constraints on the Wilson coefficients $C_{7}^{\text{eff}}$ and $C'_{7}$ using Eq.~\ref{eq:lambdagamma}.
In principle, the same method would also apply in the presence of process independent corrections to the Wilson coefficients and could also be translated in terms of $C_{7}^{\text{eff}}$ and $C'_{7}$ with theoretical input on these corrections.

These constraints could then be compared to those set by other relevant observables such as the \decay{\Bz}{\Kstarz\epem} angular observables, the time-dependent decay rate of \decay{\Bs}{\phi\gamma} decays, the \CP asymmetry in \decay{\Bz}{\Kstarz\gamma} decays or the inclusive \decay{\B}{X_s\gamma} branching fraction, which are discussed extensively in Ref.~\cite{Paul:2016urs}.
While the particular dependence of \lg on the Wilson coefficients makes this observable a priori less interesting to size non-SM effects, the statistical power of the studies shown here will compensate this limitation.
Additionally, since the dependence of \lg on the Wilson coefficients is different from that of the other observables, its measurement provides complementary information;
in particular, a measurement of \lg in \text{\BtoKpipig} decays could help break an ambiguity that arises in the determination of $\mathcal{R}e(C_{7}^\prime)$ when constraints from all radiative observables are combined assuming both Wilson coefficients to be real~\cite{Paul:2016urs}.

However, as already mentioned, theory calculations of the hadronic contributions are crucial to be able to perform this intepretation of \lg in terms of the Wilson coefficients.
Additionally, the effect of the process dependent corrections, which are disregarded at the moment, should be estimated or taken into account as nuisance parameters.

In summary, the measurement of the photon polarisation parameter through an amplitude analysis of \BtoKpipig decays is a very promising method that could exploit the large data samples available at \lhcb and Belle II in the near future.
If the current shortcomings in the interpretation are overcome, the proposed approach will allow to set very competitive and complementary new constraints on the Wilson coefficients $C_{7}^{\text{eff}}$ and $C_{7}^\prime$, and will pave the way to a new array of measurements involving decays of \bquark hadrons to three hadrons and a photon.

\section*{Acknowledgements}

We would like to thank Michael Gronau and Dan Pirjol for their guidance in understanding the details of the \decay{\B}{\Kpipi\gamma} decay, Sébastien Descotes-Genon for his valuable help in clarifying the theoretical formalism and sorting out sign discrepancies, and David Straub for the discussions and clarifications regarding the interpretation of \lg in terms of the Wilson coefficients.
Support by the Swiss National Science Fondation under contracts 166208 and 168169 is gratefully acknowledged.

\clearpage 

\begin{appendices}
\renewcommand{\thetable}{\Alph{section}.\arabic{table}}
\section{Spin factors}
\label{app:SpinFactors}

\setcounter{table}{0}

The description of the spin structure of \BtoKpipig decays is encoded in spin factors that are determined using the covariant-tensor formalism.
The spin factors are constructed such that they satisfy Lorentz invariance, angular momentum conservation, and, when applicable, parity conservation.
The three objects from which spin factors are built, namely polarisation vectors, spin projectors and angular momentum tensors, are presented briefly here. More details can be found in Refs.~\cite{PhysRev.140.B97,Chung:1997jn}.

Massive particles of mass $M$, four-momentum $p$, spin 1 and spin projection $m$ are represented in momentum space by a polarisation vector $\epsilon^{\mu}(p, m)$  that is orthogonal to the four-momentum $p$, leaving three degrees of freedom (hence three polarisation states $m = -1, 0, 1$). 
In the case of a massless particle, a particular choice of gauge is made by requiring $\epsilon^{0} = 0$, leaving only two polarisation states ($m = -1, 1$).
Spin-2 polarisation tensors are then obtained by coupling spin-1 polarisation vectors,

\begin{equation}
\epsilon^{\mu\nu}(p, m) = \sum_{m_{1}, m_{2}} \langle 1 m_{1}, 1 m_{2} | 2 m \rangle \epsilon^{\mu}(p, m_{1}) \epsilon^{\nu}(p, m_{2}),
\end{equation}

\noindent where $\langle 1 m_{1}, 1 m_{2} | 2 m \rangle$ are Clebsch-Gordon coefficients. 
By construction, the polarisation tensors satisfy the Rarita-Schwinger conditions: they are traceless, symmetric and orthogonal to $p$.

To project any tensor on the subspace spanned by a set of these polarisation tensors, operators called spin projectors are used. The spin-1 projection operator associated with a massive particle is defined as

\begin{equation}
P^{\mu\nu}_{(1)}(p) = \sum_{m} \epsilon^{\mu}(p, m)\epsilon^{*\nu}(p, m) = -g^{\mu\nu} + \dfrac{p^{\mu}p^{\nu}}{M^{2}},
\end{equation}
\noindent where $g^{\mu\nu} = \text{diag}(+1, -1, -1, -1)$ is the Minkowski metric.
The spin-2 projection operator can then be obtained from the spin-1 projection operator as

\begin{align}
P^{\mu\nu\alpha\beta}_{(2)}(p) &= \sum_{m} \epsilon^{\mu\nu}(p, m)\epsilon^{*\alpha\beta}(p, m)\\
&= \dfrac{1}{2} \left(P^{\mu\alpha}_{(1)}(p)P^{\nu\beta}_{(1)}(p) + P^{\mu\beta}_{(1)}(p) P^{\nu\alpha}_{(1)}(p)\right) - \dfrac{1}{3}P^{\mu\nu}_{(1)}(p)P^{\alpha\beta}_{(1)}(p).
\end{align}

Finally, the angular momentum tensor that describes a two-particle state of pure angular momentum $L$ is obtained from the total four-momentum $p_{R} = p_{1} + p_{2}$ and the relative four-momentum $q_{R} = p_{1} - p_{2}$, where $p_{1}$ and $p_{2}$ are the final-state four-momenta.
The angular momentum tensor is built by projecting the rank-$L$ tensor of relative momenta $q_{R}^{\nu_{1}}q_{R}^{\nu_{2}}...q_{R}^{\nu_{L}}$ on the spin-$L$ subspace

\begin{equation}
L_{(L)\mu_{1}\mu_{2}...\mu_{L}} (p_{R}, q_{R})= (-1)^{L}P_{(L)\mu_{1}\mu_{2}...\mu_{L}\nu_{1}\nu_{2}...\nu_{L}} (p_{R}) q_{R}^{\nu_{1}}q_{R}^{\nu_{2}}...q_{R}^{\nu_{L}}\,,
\end{equation}
where the spin projection tensor reduces the number of degrees of freedom from $4^{L}$ to $2L+1$.

The spin factors considered in the present study are those that describe decays of the type $B \rightarrow \gamma R_{i}, R_{i} \rightarrow P_1 R_{j}, R_{j} \rightarrow P_2 P_3 $, where $P_1$, $P_2$ and $P_3$ are the pseudoscalar particles corresponding to the final-state kaon and pions.
The spin projection of the photon is denoted~$m_{\gamma}$.
A right-handed photon corresponds to $m_{\gamma} = +1$ and a left-handed photon to $m_{\gamma} = -1$.
In general, the spin factor for such a decay can be written as a sum over the allowed spin projections of the resonances $R_i$ and $R_j$

\begin{equation}
\mathcal{S}^{ij,m_{\gamma}} = \sum_{m_i,m_j} \langle P_2 P_3|\mathcal{M}|R_{j}(m_j)\rangle  \langle R_{j}(m_j) P_1 |\mathcal{M}|R_{i}(m_i)\rangle \langle R_{i}(m_i) \gamma(m_{\gamma}) |\mathcal{M}|B \rangle,
\end{equation}
where $\mathcal{M}$ is the matrix element of the relevant decay. 
Each of the terms associated with a two-body process $R \rightarrow AB$ with a spin-orbit configuration $(L_{AB}, S_{AB})$ is expressed as

\begin{equation}
\left\langle AB, L_{AB}, S_{AB} \right| \mathcal{M} \left| R \right\rangle =  \varepsilon_{(S_{R})} (R) X (S_{R}, L_{AB}, S_{AB}) L_{(L_{AB})} (R) \Phi_{(S_{AB})}\,,
\end{equation}
where

\begin{equation}
\Phi_{(S_{AB})} =  P_{(S_{AB})} (R) X (S_{AB}, S_{A}, S_{B}) \varepsilon^{*}_{(S_{A})} (A) \varepsilon^{*}_{(S_{B})} (B)\,.
\end{equation}

The term $\varepsilon_{(S_{R})} (R) $ is a polarisation tensor assigned to the decaying particle and $\varepsilon^{*}_{(S_{A})} (A)$ and $\varepsilon^{*}_{(S_{B})} (B)$ are conjugated polarisation tensors assigned to the children particles.
The spin projector $P_{(S_{AB})} (R)$ and the angular momentum tensor $L_{(L_{AB})} (R)$ describe the spin and angular momentum coupling, respectively.
All tensors are contracted to give a scalar, requiring in some cases the inclusion of the tensor $\varepsilon_{\alpha\beta\gamma\delta}u^{\delta}_{R}$ through

\begin{equation}
X(j_{a}, j_{b}, j_{c}) = 
  \begin{cases}
    1 & \text{for } j_{a} + j_{b} + j_{c} \text{ even}\,, \\
    \varepsilon_{\alpha\beta\gamma\delta}u^{\delta}_{R} & \text{for } j_{a} + j_{b} + j_{c} \text{ odd}\,, 
  \end{cases}
\end{equation}
where $u_{R}$ is the momentum of resonance $R$ divided by its invariant mass, $u_{R} = p_{R}/M_{R}$.

To obtain the spin factor associated with a given decay chain, the various two-body processes are combined and all the allowed spin projections that are not distinguishable are summed.
This implies that the sum is performed on all the spin projections of the hadrons present in the decay chains, but not on the spin projections of the photon.
In the end, the expression of the spin factor only depends on the spin projection of the photon and on the spin-parity of the resonances $R_i$ and $R_j$.
The spin factors obtained for the decay chains used in this paper  are shown in Table~\ref{tab:spinFactors}.

\begin{table}[hb]
    \caption{Spin factors for different decay chains leading to $B \rightarrow P_{1} P_{2} P_{3} \gamma$.
    The letters S, P, V, A refer to scalar, pseudoscalar, vector and axial-vector particles, respectively. $T_{+}$ and $T_{-}$ are tensor particles with positive and negative parity, respectively.
    By default, the lowest total angular momentum $L_{AB}$ accessible in each of the two-body decays is used.
    The symbol $[D]$ refers to decay chains where $L_{AB}$ is set to 2.}
    \begin{center}
    \resizebox{\textwidth}{!}{%
    \begin{tabular}{ll}
        \toprule
        Decay chain & Spin factor \\
        \midrule
        $B \rightarrow A \gamma, A \rightarrow V P_{1}, V \rightarrow P_{2} P_{3}$ & $\epsilon^{*}_{\alpha}(\gamma) P^{\alpha\beta}_{(1)}(A)L_{(1)\beta}(V) $ \\ [5pt]
        $B \rightarrow A \gamma, A[D] \rightarrow V P_{1}, V \rightarrow P_{2} P_{3}$ & $\epsilon^{*}_{\alpha}(\gamma) L^{\alpha\beta}_{(2)}(A)L_{(1)\beta}(V) $ \\[5pt]
        $B \rightarrow A \gamma, A \rightarrow S P_{1}, S \rightarrow P_{2} P_{3}$ & $\epsilon^{*\alpha}(\gamma) L_{(1)\alpha}(A) $ \\[5pt]
        $B \rightarrow V_{1} \gamma, V_{1} \rightarrow V_{2} P_{1}, V_{2} \rightarrow P_{2} P_{3}$ & $\epsilon^{*}_{\alpha}(\gamma) P^{\alpha\kappa}_{(1)}(V_{1}) \epsilon_{\kappa\lambda\mu\nu} L^{\lambda}_{(1)}(V_{1})u^{\mu}_{V_{1}} P^{\nu\xi}_{(1)} (V_{1}) L_{(1)\xi}(V_{2}) $ \\[5pt]
        $B \rightarrow T_{-} \gamma, T_{-} \rightarrow V P_{1}, V \rightarrow P_{2} P_{3}$ & $L_{(1)\alpha}(B) \epsilon^{*}_{\beta}(\gamma) P^{\alpha\beta\lambda\mu}_{(2)}(T_{-}) L_{(1)\lambda}(T_{-}) P_{(1)\mu\nu}(T_{-}) L_{(1)}^{\nu}(V)  $ \\[5pt]
        $B \rightarrow T_{-} \gamma, T_{-} \rightarrow S P_{1}, S \rightarrow P_{2} P_{3}$ & $L_{(1)\alpha}(B) \epsilon^{*}_{\beta}(\gamma) L^{\alpha\beta}_{(2)}(T_{-})  $ \\[5pt]
        $B \rightarrow T_{+} \gamma, T_{+} \rightarrow V P_{1}, V \rightarrow P_{2} P_{3}$ & $\epsilon_{\kappa\lambda\mu\nu} u_{T_+}^{\kappa} L_{(1)\alpha}(B) \epsilon^{*}_{\beta}(\gamma) P^{\alpha\beta\lambda\xi}_{(2)}(T_{+}) L^{\mu}_{(2)\xi}(T_{+}) P^{\nu\rho}_{(1)} (T_{+}) L_{(1)\rho}(V) $ \\[5pt]
        \bottomrule
        \label{tab:spinFactors}
    \end{tabular}}
    \end{center}
\end{table}

\clearpage
\section{Additional sensitivity studies}
\label{app:Sensitivity}

\setcounter{table}{0}

The tables below present results of fits performed on simulated \BptoKpipig samples generated with two amplitudes: $K_1(1270)^+ \to \Kp\rho(770)^0$ and $K_1(1270)^+  \rightarrow K^{*}(892)^0 \pi^+$.
Table~\ref{tab:fitResults_2amps_RH} lists results of fits on simulated data sets generated with different relative magnitudes and phases. 
Results of fits for models generated with various values of $\lambda_{\gamma}$ are shown for two sets of two-amplitude samples corresponding respectively to a region of high up-down asymmetry (relative phase of $-0.91$) in Table~\ref{tab:fitResults_2amps_highAud} and a region of low up-down asymmetry (relative phase of $0.82$) in Table~\ref{tab:fitResults_2amps_lowAud}.

 \begin{table}[b]
    \caption{Results of unbinned maximum likelihood fits for 100 pseudo-experiments, for simplified two-amplitude models generated with $\lambda_{\gamma}=1$, for various relative magnitudes and phases. The parameters $a$ and $\phi$ stand respectively for the relative magnitude and phase between the decay with $K_1(1270)^+ \to \Kp\rho(770)^0$ and the decay with $K_1(1270)^+  \rightarrow K^{*}(892)^0 \pi^+$. The value $\phi = -0.91$ corresponds to a region of high up-down asymmetry while the values $0.82$ and $-2.32$ correspond to a region of low up-down asymmetry. The magnitudes $a = 1.01$, $2.02$ and $3.03$ correspond to ratios of fractions between the two amplitudes of $0.62$, $2.47$ and $5.57$, respectively.\label{tab:fitResults_2amps_RH}}
    \begin{center}
 \begin{tabular}{cccccc}
        \toprule
 Parameter          & True value & Mean value         & Std deviation & $\phantom{-}\mu_{\textrm{pull}}$ & $\sigma_{\textrm{pull}}$ \\ \midrule
$a$            		    & $\phantom{-}2.02$& $\phantom{-}2.017$ & $0.03$        & $\phantom{-}0.10 \pm 0.1$0       & $1.04 \pm 0.07$ \\
$\phi$     		        & $-0.91$&$-0.909$           & $0.02$        & $-0.09 \pm 0.10$                 & $1.05 \pm 0.07$  \\
$\lambda_{\gamma}$ & $\phantom{-0.0}1$ & $\phantom{-}1.002$ & $0.04$        & $-0.14 \pm 0.10$                 & $1.09 \pm 0.07$\\ \midrule
$a$            		    &$\phantom{-}2.02$& $\phantom{-}2.020$ & $0.04$        & $\phantom{-}0.01 \pm 0.11$       & $1.14 \pm 0.07$ \\
$\phi$      		       & $\phantom{-}0.82$ & $\phantom{-}0.823$ & $0.02$        & $-0.09 \pm 0.09$                 & $0.94 \pm 0.07$ \\
$\lambda_{\gamma}$ & $\phantom{-0.0}1$ & $\phantom{-}1.001$ & $0.04$        & $-0.09 \pm 0.12$                 & $1.17 \pm 0.08$ \\ \midrule
$a$        		        & $\phantom{-}2.02$& $\phantom{-}2.021$ & $0.03$        & $-0.03 \pm 0.10$                 & $0.98 \pm 0.07$ \\
$\phi$    		         & $-2.32$& $-2.318$           & $0.02$        & $-0.11 \pm 0.10$                 & $1.07 \pm 0.07$ \\
$\lambda_{\gamma}$ & $\phantom{-0.0}1$ & $\phantom{-}1.001$ & $0.02$        & $-0.08 \pm 0.11$                 & $1.11 \pm 0.07$ \\ \midrule
$a$        		        & $\phantom{-}1.01$& $\phantom{-}1.011$ & $0.02$        & $-0.06 \pm 0.11$                 & $1.16 \pm 0.08$ \\
$\phi$    		         & $-0.91$& $-0.908$           & $0.03$        & $-0.09 \pm 0.11$                 & $1.14 \pm 0.07$ \\
$\lambda_{\gamma}$ & $\phantom{-0.0}1$& $\phantom{-}1.002$ & $0.04$        & $-0.11 \pm 0.11$                 & $1.12 \pm 0.07$ \\ \midrule
$a$        		        & $\phantom{-}3.03$& $\phantom{-}3.028$ & $0.06$        & $\phantom{-}0.06 \pm 0.10$       & $1.03 \pm 0.07$ \\
$\phi$    		         & $-0.91$& $-0.907$           & $0.03$        & $-0.09 \pm 0.10$                 & $1.02 \pm 0.07$ \\
 $\lambda_{\gamma}$ & $\phantom{-0.0}1$& $\phantom{-}1.006$ & $0.03$        & $-0.36 \pm 0.10$                 & $1.08 \pm 0.07$\\
        \bottomrule
\end{tabular}
\end{center}
 
\end{table}

\begin{table}[tb]
    \caption{Results of unbinned maximum likelihood fits for 100 generated data sets simulated according to the simplified two-amplitude model with relative magnitude and phase $(2.02, -0.91)$ corresponding to a ratio of fractions of $2.47$, and various values of $\lambda_{\gamma}$. The parameters $a$ and $\phi$ stand respectively for the relative magnitude and phase between the decay with $K_1(1270)^+ \to \Kp\rho(770)^0$ and the decay with $K_1(1270)^+  \rightarrow K^{*}(892)^0 \pi^+$.\label{tab:fitResults_2amps_highAud}} 
    \begin{center}
    \begin{tabular}{cccccc}
  \toprule
Parameter                  & True value & Mean value         & Std deviation & $\phantom{-}\mu_{\textrm{pull}}$ & $\sigma_{\textrm{pull}}$ \\
  \midrule
$a$                  		      & $\phantom{-}2.02$ & $\phantom{-}2.017$ & $0.03$        & $\phantom{-}0.10 \pm 0.10 $      & $1.04 \pm 0.07 $ \\
$\phi$             		      & $-0.91$ & $-0.909$           & $0.02$        & $-0.09 \pm 0.10$                 & $1.05 \pm 0.07  $ \\
$\lambda_{\gamma}$   & $\phantom{-0.0}1$ & $\phantom{-}1.002$ & $0.04$        & $-0.14 \pm 0.10$                 & $1.09 \pm 0.07$ \\
  \midrule
$a$               		         & $\phantom{-}2.02$& $\phantom{-}2.017$ & $0.03$        & $\phantom{-}0.10 \pm 0.10$       & $ 1.03 \pm 0.07$ \\
$\phi$            		         & $-0.91$& $-0.911$           & $0.03$        & $\phantom{-}0.03 \pm 0.11$       & $ 1.14 \pm 0.08$ \\
$\lambda_{\gamma}$  & $0.875$& $\phantom{-}0.873$ & $0.04$        & $\phantom{-}0.03 \pm 0.10$       & $1.21 \pm 0.08$ \\
  \midrule
$a$                     	   & $\phantom{-}2.02$& $\phantom{-}2.019$ & $0.03$        & $\phantom{-}0.06 \pm 0.10$       & $1.05 \pm 0.07$ \\
$\phi$                     		& $-0.91$& $-0.911$           & $0.02$        & $\phantom{-}0.03 \pm 0.10$       & $  0.99 \pm 0.07 $ \\
$\lambda_{\gamma}$ & $\phantom{-}0.75$& $\phantom{-}0.751$ & $0.04$        & $-0.05 \pm 0.11$                 & $ 1.25 \pm 0.08$ \\
\bottomrule
 \end{tabular}
 \end{center}
   
\end{table}

\begin{table}[tb]
    \caption{Results of unbinned maximum likelihood fits for 100 pseudo-experiments, simulated according to the simplified two-amplitude model with relative magnitude and phase $(2.02, 0.82)$ corresponding to a ratio of fractions of $2.47$, and various values of $\lambda_{\gamma}$. The parameters $a$ and $\phi$ stand respectively for the relative magnitude and phase between the decay with $K_1(1270)^+ \to \Kp\rho(770)^0$ and the decay with $K_1(1270)^+  \rightarrow K^{*}(892)^0 \pi^+$.\label{tab:fitResults_2amps_lowAud}} 
    \begin{center}
    \begin{tabular}{cccccc}
  \toprule
Parameter          & True value & Mean value & Std deviation & $\phantom{-}\mu_{\textrm{pull}}$ & $\sigma_{\textrm{pull}}$ \\ \midrule
$a$                & $\phantom{-}2.02$& $2.020$    & $0.04$        & $\phantom{-}0.01 \pm 0.11$       & $1.14 \pm 0.07$ \\
$\phi$             & $\phantom{-}0.82$& $0.823$    & $0.02$        & $-0.09 \pm 0.09$                 & $0.94 \pm 0.07  $ \\
$\lambda_{\gamma}$ & $\phantom{-0.0}1$& $1.001$    & $0.04$        & $-0.09 \pm 0.12$                 & $1.17 \pm 0.08$ \\ \midrule
$a$                & $\phantom{-}2.02$& $2.023$    & $0.04$        & $-0.06 \pm 0.11$                 & $1.17 \pm 0.07$ \\
$\phi$             & $\phantom{-}0.82$& $0.823$    & $0.03$        & $-0.13 \pm 0.09$                 & $0.97 \pm 0.06$ \\
$\lambda_{\gamma}$ & $0.875$& $0.870$    & $0.04$        & $\phantom{-}0.11 \pm 0.11$       & $1.17 \pm 0.08$ \\ \midrule
$a$                & $\phantom{-}2.02$& $2.022$    & $0.04$        & $-0.03 \pm 0.09$                 & $1.03 \pm 0.07 $ \\
$\phi$             & $\phantom{-}0.82$& $0.822$    & $0.03$        & $-0.07 \pm 0.09$                 & $0.92 \pm 0.06 $ \\
$\lambda_{\gamma}$ & $\phantom{-}0.75$& $0.741$    & $0.04$        & $\phantom{-}0.20 \pm 0.09$       & $1.03 \pm 0.07$ \\
\bottomrule
 \end{tabular} 
 \end{center}

\end{table}

\clearpage

\clearpage
\end{appendices}

\addcontentsline{toc}{section}{References}
\printbibliography

\end{document}